\documentclass[a4paper,11pt]{article}
\pdfoutput=1 

\usepackage{jcappub} 

\usepackage[T1]{fontenc} 

\usepackage{color}

\usepackage{graphicx}
\usepackage{dcolumn}
\usepackage{bm}
\usepackage{xspace}
\usepackage{array}
\usepackage{xfrac}
\usepackage{dcolumn}
\usepackage[normalem]{ulem}
\newcolumntype{L}[1]{>{\raggedright\let\newline\\\arraybackslash\hspace{0pt}}m{#1}}
\newcolumntype{C}[1]{>{\centering\let\newline\\\arraybackslash\hspace{0pt}}m{#1}}
\newcolumntype{R}[1]{>{\raggedleft\let\newline\\\arraybackslash\hspace{0pt}}m{#1}}
\newcolumntype{d}[1]{D{.}{.}{#1}}
\newcommand{\comment}[1]{} 

\newcommand{\znbb}      {$0\nu\beta\beta$\xspace}
\newcommand{\tnbb}      {$2\nu\beta\beta$\xspace}
\newcommand{\otsx}      {$^{136}\mathrm{Xe}$\xspace}
\newcommand{\dd}        {\mathrm{d}}

\title{\boldmath Cosmogenic Backgrounds to $0\nu\beta\beta$ in EXO-200}

\author[a,1]{J.B.~Albert,\note{Corresponding author.}}
\author[b,2]{D.J.~Auty,\note{Now at University of Alberta, Edmonton, Alberta, Canada}}
\author[c]{P.S.~Barbeau,}
\author[d]{D.~Beck,}
\author[e]{V.~Belov,}
\author[f]{M.~Breidenbach,}
\author[g,h]{T.~Brunner}
\author[e]{A.~Burenkov,}
\author[i]{G.F.~Cao,}
\author[j]{C.~Chambers,}
\author[k,3]{B.~Cleveland,\note{Also SNOLAB, Sudbury ON, Canada}}
\author[d]{M.~Coon,}
\author[j]{A.~Craycraft,}
\author[f]{T.~Daniels,}
\author[e]{M.~Danilov,}
\author[a]{S.J.~Daugherty,}
\author[f]{J.~Davis,}
\author[l]{S.~Delaquis,}
\author[k]{A.~Der Mesrobian-Kabakian,}
\author[m]{R.~DeVoe,}
\author[b]{T.~Didberidze,}
\author[h]{J.~Dilling,}
\author[e]{A.~Dolgolenko,}
\author[n]{M.J.~Dolinski,}
\author[o]{M.~Dunford,}
\author[j]{W.~Fairbank Jr.,}
\author[k]{J.~Farine,}
\author[p]{W.~Feldmeier,}
\author[q]{S.~Feyzbakhsh,}
\author[p]{P.~Fierlinger,}
\author[m]{D.~Fudenberg,}
\author[o,h]{R.~Gornea,}
\author[o]{K.~Graham,}
\author[m]{G.~Gratta,}
\author[r]{C.~Hall,}
\author[f]{S.~Herrin,}
\author[b]{M.~Hughes,}
\author[m]{M.J.~Jewell,}
\author[f]{A.~Johnson,}
\author[a]{T.N.~Johnson,}
\author[q]{S.~Johnston,}
\author[e]{A.~Karelin,}
\author[a]{L.J.~Kaufman,}
\author[o]{R.~Killick,}
\author[o]{T.~Koffas,}
\author[m]{S.~Kravitz,}
\author[h]{R.~Kr\"{u}cken,}
\author[e]{A.~Kuchenkov,}
\author[s]{K.S.~Kumar,}
\author[t]{D.S.~Leonard,}
\author[o]{C.~Licciardi,}
\author[n]{Y.H.~Lin,}
\author[d,4]{J.~Ling,\note{Now at Sun Yat-Sen University, Guangzhou, China}}
\author[u]{R.~MacLellan,}
\author[p]{M.G.~Marino,}
\author[f]{B.~Mong,}
\author[m]{D.~Moore,}
\author[s]{O.~Njoya,}
\author[v]{R.~Nelson,}
\author[f]{A.~Odian,}
\author[m]{I.~Ostrovskiy,}
\author[b]{A.~Piepke,}
\author[q]{A.~Pocar,}
\author[f]{C.Y.~Prescott,}
\author[h]{F.~Reti\`{e}re,}
\author[f]{P.C.~Rowson,}
\author[f]{J.J.~Russell,}
\author[m]{A.~Schubert,}
\author[o,h]{D.~Sinclair,}
\author[n]{E.~Smith,}
\author[e]{V.~Stekhanov,}
\author[s]{M.~Tarka,}
\author[l,5]{T.~Tolba,\note{Now at Institute of High Energy Physics, Beijing, China}}
\author[b]{R.~Tsang,}
\author[m]{K.~Twelker,}
\author[l]{J.-L.~Vuilleumier,}
\author[f]{A.~Waite,}
\author[d]{J.~Walton,}
\author[j]{T.~Walton,}
\author[m]{M.~Weber,}
\author[i]{L.J.~Wen,}
\author[k]{U.~Wichoski,}
\author[v]{J.~Wood,}
\author[d]{L.~Yang,}
\author[n]{Y.-R.~Yen,}
\author[e]{O.Ya.~Zeldovich}

\affiliation[a]{Physics Department and CEEM, Indiana University, Bloomington, Indiana 47405, USA}
\affiliation[b]{Department of Physics and Astronomy, University of Alabama, Tuscaloosa, Alabama 35487, USA}
\affiliation[c]{Department of Physics, Duke University, and Triangle Universities Nuclear Laboratory (TUNL), Durham, North Carolina 27708, USA}
\affiliation[d]{Physics Department, University of Illinois, Urbana-Champaign, Illinois 61801, USA}
\affiliation[e]{Institute for Theoretical and Experimental Physics, Moscow, Russia}
\affiliation[f]{SLAC National Accelerator Laboratory, Menlo Park, California 94025, USA}
\affiliation[g]{Physics Department, McGill University, Montreal, Quebec H3A 2T8, Canada}
\affiliation[h]{TRIUMF, Vancouver, British Columbia V6T 2A3, Canada}
\affiliation[i]{Institute of High Energy Physics, Beijing, China}
\affiliation[j]{Physics Department, Colorado State University, Fort Collins, Colorado 80523, USA}
\affiliation[k]{Department of Physics, Laurentian University, Sudbury, Ontario P3E 2C6, Canada}
\affiliation[l]{LHEP, Albert Einstein Center, University of Bern, Bern, Switzerland}
\affiliation[m]{Physics Department, Stanford University, Stanford, California 94305, USA}
\affiliation[n]{Department of Physics, Drexel University, Philadelphia, Pennsylvania 19104, USA}
\affiliation[o]{Physics Department, Carleton University, Ottawa, Ontario K1S 5B6, Canada}
\affiliation[p]{Technische Universit\"at M\"unchen, Physikdepartment and Excellence Cluster Universe, Garching 85748, Germany}
\affiliation[q]{Amherst Center for Fundamental Interactions and Physics Department, University of Massachusetts, Amherst, MA 01003, USA}
\affiliation[r]{Physics Department, University of Maryland, College Park, Maryland 20742, USA}
\affiliation[s]{Department of Physics and Astronomy, Stony Brook University, SUNY, Stony Brook, New York 11794, USA}
\affiliation[t]{IBS Center for Underground Physics, Daejeon, Korea}
\affiliation[u]{Physics Department, University of South Dakota, Vermillion, South Dakota 57069, USA}
\affiliation[v]{Waste Isolation Pilot Plant, Carlsbad, New Mexico 88220, USA}

\emailAdd{joalbert@indiana.edu}

\abstract{As neutrinoless double-beta decay experiments become more sensitive and intrinsic 
radioactivity in detector materials is reduced, previously minor contributions to the background
must be understood and eliminated.  With this in mind, cosmogenic backgrounds have
been studied with the EXO-200 experiment.  Using the EXO-200 TPC, 
the muon flux (through a flat horizontal surface) underground at the Waste Isolation Pilot Plant (WIPP) has been measured 
to be
$\Phi = 4.07\pm0.14~\mathrm{(sys)}\pm0.03~\mathrm{(stat)}\times 10^{-7} \mathrm{cm^{-2}~s^{-1}}$,
with a vertical intensity of 
$I_{\mathrm{v}} = 2.97^{+0.14}_{-0.13}~\mathrm{(sys)}\pm0.02~\mathrm{(stat)}\times 10^{-7} \mathrm{cm^{-2}~s^{-1}~sr^{-1}}$.  
Simulations of muon-induced backgrounds identified several potential cosmogenic radionuclides,
though only $^{137}\mathrm{Xe}$ is a significant background for the \otsx \znbb search with EXO-200.
Muon-induced 
neutron backgrounds were measured using $\gamma$-rays from neutron capture on the 
detector materials.  This provided a measurement of $^{137}\mathrm{Xe}$ yield, and
a test of the accuracy of the neutron production and transport simulations.  The independently measured rates
of $^{136}\mathrm{Xe}$ neutron capture and of $^{137}\mathrm{Xe}$ decay
agree within uncertainties. Geant4 and FLUKA simulations were performed to estimate neutron capture rates,
and these estimates agreed to within $\sim$40\% or better with measurements.
The ability to identify $^{136}\mathrm{Xe}(n,\gamma)$ events will allow for rejection of
$^{137}\mathrm{Xe}$ backgrounds in future \znbb analyses.}

\begin{document}
\maketitle
\flushbottom

\section{\label{sec:intro}Introduction}

While neutrino oscillation experiments have demonstrated that neutrinos are massive
particles, their behavior under CP conjugation (Dirac or Majorana) and the absolute value 
of their masses are still
unknown.  The observation of neutrinoless double-beta decay (\znbb), a hypothetical 
lepton-number-violating decay mode, would demonstrate that neutrinos are Majorana 
particles and could determine the absolute neutrino mass scale.
Certain even-even nuclei, such as \otsx and $^{76}\mathrm{Ge}$, are stable against
ordinary $\beta$ decay but have been observed to undergo two-neutrino double-beta decay (\tnbb) 
and are candidates for \znbb searches.  Current limits on the half
lives for the \znbb mode of \otsx and $^{76}\mathrm{Ge}$ have reached the $10^{25}$
year level~\cite{Gando:2012zm,Albert:2014awa,Agostini:2013mzu}.
 Understanding and mitigation of all backgrounds is critical for further improvement.

Backgrounds to \znbb experiments can be divided into two main categories.  First are
those produced by long-lived radionuclides  (typically U and Th).  These can
be mitigated through shielding, removal of surface activity, 
and careful materials selection during detector construction and design.
Second are backgrounds induced by cosmic rays and their products.  
These cannot be completely shielded against, but certain techniques, as well as the use of
deep underground laboratories, can mitigate their
impact on experimental sensitivity.  This paper reports a study of cosmic-ray-induced backgrounds 
and, in particular, neutron capture, with EXO-200.
Monte Carlo (MC) simulations and data taken with the detector were utilized
for this purpose.

\subsection{The EXO-200 detector}
\begin{figure}[htb]
	\begin{center}
	\includegraphics[keepaspectratio=true,width=\textwidth, trim = 0mm 0mm 0mm 0mm]
	{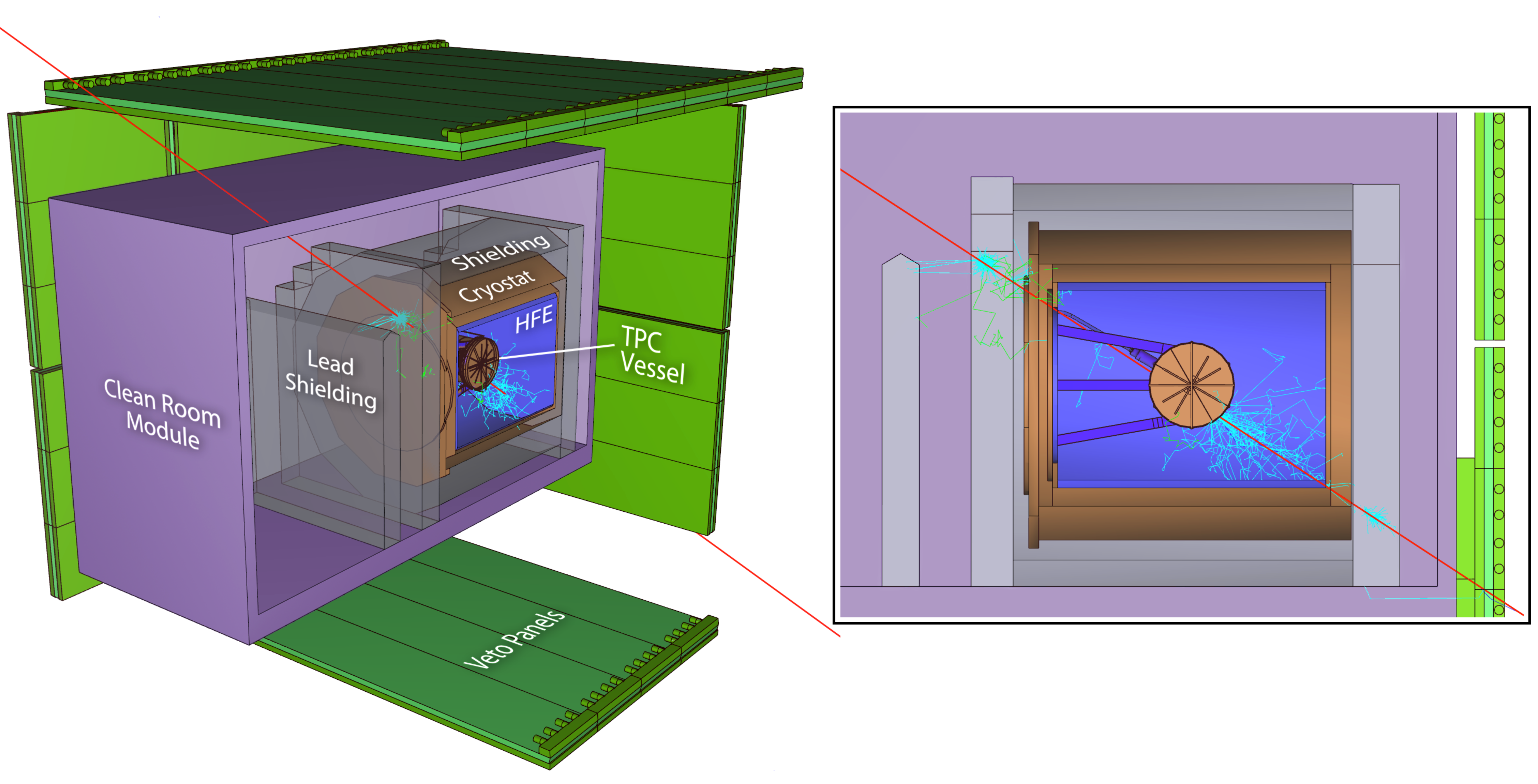}
	\end{center}
    \caption[EXO-200 rendering]
    {The EXO-200 cleanroom with various components labeled.  The excavated cavern (not shown) extends $\sim$60~cm 
    above the top veto panels and $\sim$25~cm below the bottom veto panels.
    The rendering was produced using our detector model with Geant4. A muon track
    (red) is included, along with photon (cyan) and neutron (green) tracks produced in the muon
    shower.  An alternate view (in box on right) shows the shower more 
    clearly.}
     \label{fig:exo200render}
\end{figure}

While a more complete description of the EXO-200 detector can be found elsewhere~\cite{Auger:2012gs},
the relevant features are discussed here.  EXO-200 utilizes a time
projection chamber (TPC) containing liquid xenon (LXe) enriched to 80.6\% \otsx 
($Q_{\beta\beta} = 2457.83 \pm 0.37~\mathrm{keV}$~\cite{Redshaw:2007un}), with 19.1\% $^{134}\mathrm{Xe}$.
The TPC has two drift regions, each with a signal readout, separated by a central cathode. 
The vessel is roughly 40~cm in diameter and 44~cm in length.  
Crossed wire planes at
either end of the TPC provide induction and charge collection signals.
Behind the wires, arrays of
avalanche photodiodes (APDs) collect scintillation light.  
The collection and induction channels are known as the U-wires and V-wires, respectively.
A teflon reflector surface is positioned just inside of the copper field rings.
Signals from the wireplanes and APDs are digitized at 1~MS/s and saved in ``frames''
of $\pm$1~ms around a data trigger.  The TPC is surrounded by a 
vacuum-insulated copper cryostat filled with 
HFE-7000~($\mathrm{C_{3}F_{7}OCH_{3}}$~\cite{3m}) cryofluid, providing at least
$50$~cm shielding in any direction.  The cryostat is surrounded by a lead shield
of $\sim$25~cm thickness.  The entire apparatus is located in a clean room $\sim$655~m
underground at the Waste Isolation Pilot Plant (WIPP) near Carlsbad, NM.  A rendering of
the detector and shielding can be found in figure~\ref{fig:exo200render}.  This rendering
includes a muon track passing through the TPC and all the muon secondary tracks.

EXO-200 has an active muon veto system.  Scintillator panels, 5~cm-thick, are located on
the outside of the cleanroom on 4 sides, providing $96.35\pm0.11\%$
efficiency for tagging muons passing through the TPC.  This allows tagging of prompt
backgrounds associated with muons, such as neutron capture $\gamma$s and muon bremsstrahlung.
The panels, refurbished from the concluded KARMEN experiment~\cite{Gemmeke:1990ix}, 
also feature a 4~cm-thick polyethylene backing, loaded with 5\% boron by 
mass.  This provides structural support and some absorption of neutrons, though 
simulations indicate that the boron has little
effect on cosmogenic neutron rates inside the cryostat. 

The \znbb signals
are distinguished from backgrounds through several active background suppression methods: 
scintillation-ionization ratio, energy, cluster multiplicity, 
and event position.
The energy of interactions in EXO-200 is computed as a linear combination of collected
charge and scintillation light.
With an energy resolution of 1.53\% at the 
\znbb Q-value~\cite{Albert:2014awa}, most radioactivity-induced backgrounds, including \tnbb, 
can be actively suppressed using these analysis tools.
Signals from $\beta$ decays and $\gamma$-scattering are easily distinguished 
from $\alpha$~decays in the xenon by the ratio of charge and scintillation light 
collected~\cite{PhysRevC.92.045504}.
In \znbb, the sum kinetic energy of the emitted electrons equals the Q-value of
the decay.  
While $\beta$ decays tend to
deposit their charge in a small volume of the detector, $\gamma$s of similar energy
will Compton scatter and typically deposit energy in two or more positions (``clusters'').  
Charge clusters spatially separated by $\sim$1~cm or more can be resolved, allowing 
classification of events as single-site (SS), characteristic of \tnbb and \znbb, or
multi-site (MS), characteristic of $\gamma$-induced signals.
Finally, the xenon, a high-Z material, is self-shielding to $\gamma$ rays and is kept extremely
pure. Nearly all $\gamma$ backgrounds originate from outside the active xenon
region \cite{Albert:2015nta}.  The attenuation of external backgrounds in the
LXe can be measured and fit to, providing further background 
identification~\cite{Albert:2013gpz}.  The signal for \znbb is thus a set of SS events at the
Q-value energy with the appropriate scintillation-ionization ratio which are 
uniformly distributed throughout the detector.  
The MS event set and SS events at other energies help characterize the composition
of the remaining background.

\subsection{Cosmogenic backgrounds}
Cosmogenic muons may produce fast neutrons and unstable nuclides in underground detectors and shielding
through several mechanisms, including muon spallation (virtual photon nuclear disintegration),
muon elastic scattering on neutrons, photonuclear reactions from electromagnetic showers,
nuclear capture of stopped muons, 
and secondary neutron reactions.
These secondary neutron interactions, specifically neutron capture,
are of particular interest to low-background studies with EXO-200.

Neutron capture may produce two signals in EXO-200.  First, prompt $\gamma$s
from nuclear de-excitation will be produced immediately after neutron capture.  This prompt
signal usually occurs in coincidence with triggers from the muon veto panels, and is thus rejected
from the data set for the \znbb searches.  Later, 
decay radiation from a generated radionuclide may produce delayed signals.  
The most prominent background contributor produced in the bulk of LXe is muon-induced 
$^{137}\mathrm{Xe}$ $\beta$ decay, with a half-life of 3.82~minutes \cite{Carlson1969267}.  
Produced through $^{136}\mathrm{Xe}(n,\gamma)$, this $\beta$-emitter (creating coincident $\gamma$s 
only $33\pm3\%$ of the time~\cite{PhysRev.136.B365}) typically passes the SS/MS rejection,
and has a near uniform position distribution (similar to that of \tnbb and \znbb).
The Q-value for this decay is $4173\pm7$~keV \cite{Browne20072173}, so the $\beta$ spectrum overlaps the 
Q-value for \znbb.  
Indeed, in the most recent EXO-200
\znbb search \cite{Albert:2014awa}, $^{137}\mathrm{Xe}$ $\beta$~decay was fitted to contribute 7.0 counts to the SS
region of interest ($\pm2~\sigma$~energy window), out of a total 
$31.1\pm1.8\mathrm{(stat.)\pm3.3(sys.)}$~counts, during a $^{\rm enr}$Xe exposure of
$123.7\; \mathrm{kg\cdot yr}$.

Understanding and mitigating this background is important for improving \otsx \znbb analyses
and for designing new experiments.
Detection of the nuclear de-excitation that occurs after
a neutron capture can be used to tag the production rates of this and other cosmogenic
nuclides.
 We searched for neutron capture 
signals in EXO-200 TPC data coincident with
muon veto panel triggers.  This veto-tagged data allows
us to search for and measure production of cosmogenic nuclides by neutron capture
independently from the fit of low-background data (the dataset with vetoes applied,
used for the \znbb analysis~\cite{Albert:2014awa}).

Three separate but related studies are reported here.
First, section~\ref{sec:mu_flux} describes a measurement of the muon flux at WIPP, using data
from muons passing through the EXO-200 TPC.  The muon flux is needed to normalize simulations
of cosmogenic backgrounds.  Next, section~\ref{sec:nu_sim} describes
simulations of cosmic-ray muons and their products at EXO-200.  Cosmogenic nuclide 
production rates are derived from these simulations and measured muon flux, providing a
catalog of potential backgrounds to \znbb.
Finally, section~\ref{sec:veto_tagged_ana} describes the study of veto-tagged data to 
measure neutron capture rates on the EXO-200 detector and shielding materials.  These
measurements are used to validate the simulation results from section~\ref{sec:nu_sim} and
to demonstrate understanding of EXO-200 cosmogenic backgrounds.

\section{\label{sec:mu_flux}Muon flux}
\subsection{\label{sec:muon_spectrum}Muon spectrum generation}
The EXO-200 experiment can detect muons both using the veto scintillator panels and the
TPC.  While more muon events are available from the scintillator panels, the analysis is
not statistically limited and studying muons passing through the TPC allows for lower 
systematic uncertainty and a better determination of the muon angular distribution.
Owing to the crossed-wire nature of the charge measuring planes,
only muons traveling at certain angles relative to the TPC
are properly reconstructed with high efficiency.  Hence, some knowledge of the underground
muon angular distribution is required to determine the overall muon acceptance.

Simulations of cosmic-ray-induced muons 
were performed using the 
Geant4-based~\cite{Allison:2006ve, Agostinelli:2002hh}
EXO-200 MC simulation software (``EXOsim''). A detailed description of this software, 
along with validation studies, can be found
elsewhere~\cite{Albert:2013gpz}.  
The original simulation geometry was expanded with the addition of the muon veto panels and
surrounding salt (the main component of the WIPP cavern walls).  The muons were generated at a height of 3.5~m above the TPC center,
from a horizontal circular plane 10.5~m in radius, in the salt above the experiment.  
This generation plane covers a solid angle of $1.37\pi$~sr around the vertical direction, 
allowing 99.43\% of the expected muon flux passing through the TPC to be simulated.
The muons were generated with a charge ratio of
$N(\mu^{+})/N(\mu^{-}) = 1.3$~\cite{Agashe:2014kda}.  We found our data was well described by
a phenomenological muon angular distribution from Miyake \cite{Miyake:1973qk}.  
This angular distribution is given by
\begin{equation}
I\left(h,\theta\right) = 
I\left(h, 0\right) \left(\cos\theta\right)^{1.53} e^{\alpha\left(\sec\theta-1\right)h}
\label{eqn:miyake_ang}
\end{equation}
where $h=1050$~meters water equivalent (mwe), named ``vertical depth'' in the Miyake report, is treated
as an empirical parameter rather than the physical depth, $\theta$ is the angle with
respect to vertical, and $\alpha = -8.0\times10^{-4}~\mathrm{(mwe)}^{-1}$.

The muon energy spectrum was based on the energy distribution at the 
surface \cite{Agashe:2014kda},
\begin{equation}
\frac{\dd N}{\dd E_{0}} \approx \frac{0.14 E_{0}^{-2.7}}{\mathrm{cm^{2}~s~sr~GeV}}
\left(\frac{1}{1+\frac{1.1 E_{0} \cos\theta}{\mathrm{115~GeV}}} 
+ \frac{0.054}{1+\frac{1.1 E_{0} \cos\theta}{\mathrm{850~GeV}}} \right).
\label{eqn:dnde0}
\end{equation}
The underground energy $E$ is related to the surface energy $E_{0}$ at slant depth $X$ 
(straight-line distance traveled by a muon for a particular
angle) using: 
\begin{equation}
E_{0} = E e^{b X} + \epsilon_{\mu}\left(e^{b X}-1\right)
\label{eqn:e0sub}
\end{equation}
with $b = 3.5\times10^{-4} (\mathrm{mwe})^{-1}$ \cite{Agashe:2014kda} and
$\epsilon_{\mu} = 693$ GeV \cite{Mei:2005gm}.  For our flat overburden site, we used a vertical depth of 
1600~mwe~to compute the slant depth, based on this measurement~(see section~\ref{sec:muon_measurement})
and others~\cite{Esch:2004zj} of the total flux.  The deviation between this value and the 
final depth reported in section~\ref{sec:muon_measurement} does not significantly affect 
the results of this analysis as the resulting difference in the muon energy distribution 
has a negligible impact on the detection efficiency.
The apparent mismatch of the overburden
and vertical depth parameter for the angular distribution will be discussed in section~\ref{sec:muon_measurement}.

For the purpose of muon flux determination, a targeting method was used to avoid 
simulation of muons not directed near the TPC, so only
0.323\% of muons passing through the generation plane were fully simulated.
The targeting algorithm generated all muons which would, based on initial direction,
pass through a cylinder of double the length and radius of the TPC.  

\subsection{\label{sec:muon_efficiency} Muon reconstruction and efficiency}

\begin{figure}[htb]\begin{center}
    \includegraphics[width=0.990\textwidth]{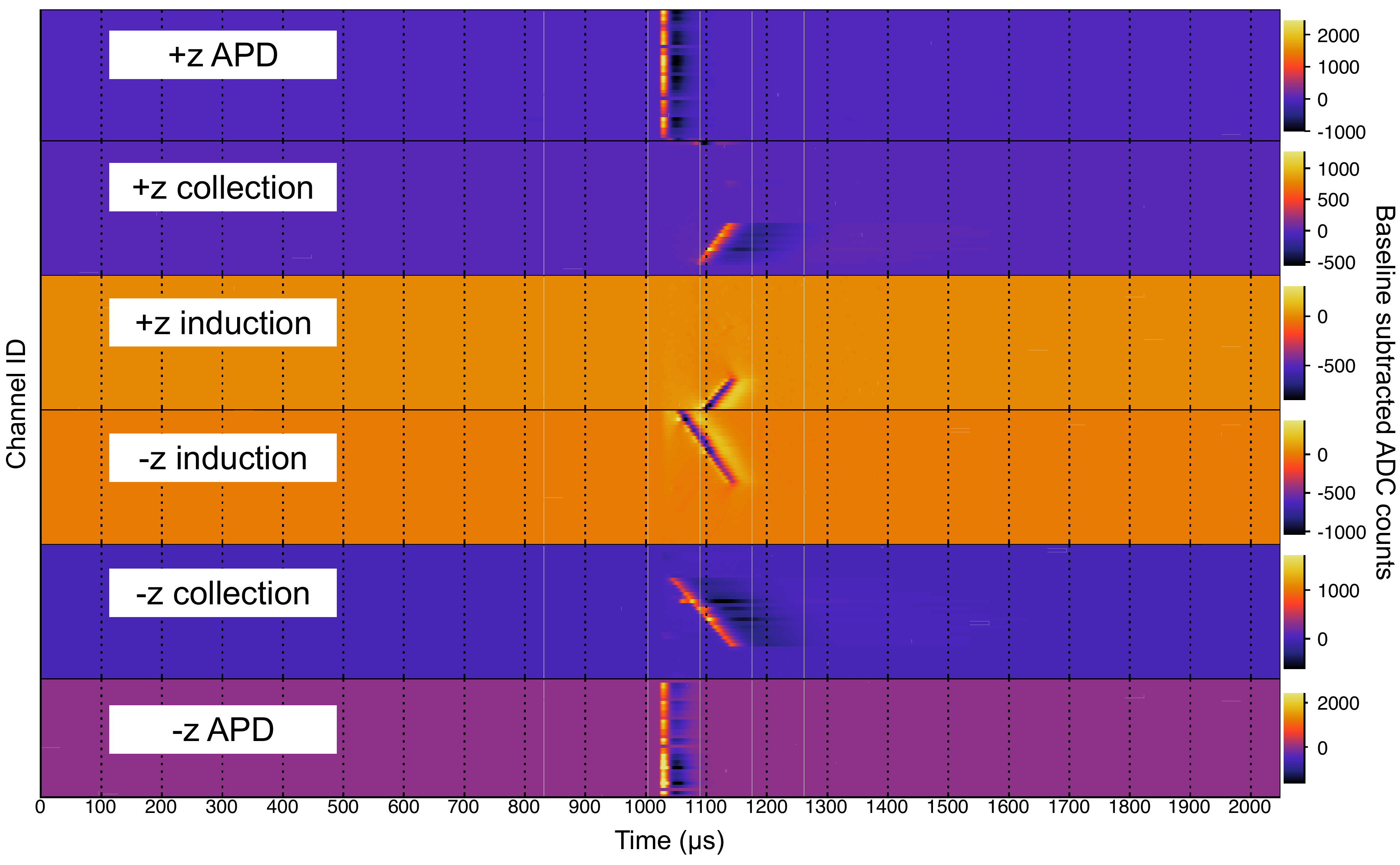}
    \caption[Muon trace]{
    A typical muon track shown on U- and V-wires and APDs, taken from EXO-200 data.  The color
    scale, indicating signal amplitude, is in baseline subtracted ADC units. There are 38 channels in each of the U- and
    V-wire planes, and 37 channels in each APD plane.  The slope between the charge collection
    and induction channels and the time coordinate is used to reconstruct the direction of the muon.  
    Because of the almost isotropic nature of light propagation inside the detector, the 
    APD light pattern indicates only the time of scintillation and total light yield, 
    and cannot help with direction reconstruction.   In 
    this particular event, the muon passes through the central cathode, so a charge signal is seen
    in both TPC halves.}
     \label{fig:HoughfromMitchell}\end{center}
\end{figure}

In EXO-200, the muon track finding algorithm is substantially different from that used to
reconstruct \znbb decays and radioactivity-induced events.  A typical 
EXO-200 TPC muon event is shown in
 figure~\ref{fig:HoughfromMitchell}.  Reconstruction requires an APD sum signal of 
at least 10000 ADC units, rejecting lower energy $\beta$s and $\gamma$s.  Muon track
signals are identified by searching the waveforms for an increase
(decrease) of 80 counts beyond the baseline on the collection (induction) wires and finding
the peak in time, here called a ``hotspot''.  
The collection of these hotspots in channel-time space is processed with a Hough 
transform~\cite{Duda:1972:UHT:361237.361242} to identify the
straight muon track and to derive the muon direction in space.  
At least 3 hotspots are required on U-wires and at least 3 on the V-wires.
Additional cuts are 
applied to reject poorly fit muon tracks and noise events.

Reconstructed positions and directions are based on a right-handed Cartesian coordinate system, 
with origin at the TPC center, and with
the $z$-axis along the TPC axis of symmetry and drift direction.
Azimuthal and polar angles ($\phi$ and $\theta$,
respectively) are measured with respect to the vertical axis, $y$, with $\theta=0$ defined
as straight down ($-y$), and $\left(\theta,\phi\right) = \left(\sfrac{\pi}{2},0\right)$ parallel to
$+z$.
The fit is applied to both halves of the TPC and, if both
halves have a well-reconstructed track, the reconstructed direction from the TPC half with
more hotspots on the track is selected.
Comparing the directions of reconstructed tracks in events where the muon passes through
both TPC halves has shown that this method works well.  While small improvements may be
possible through a combined fit of both TPC halves, we do not expect such a technique to
improve statistical or systematic uncertainties in any significant way.

\begin{figure}\begin{center}
\includegraphics[width=0.49\textwidth]{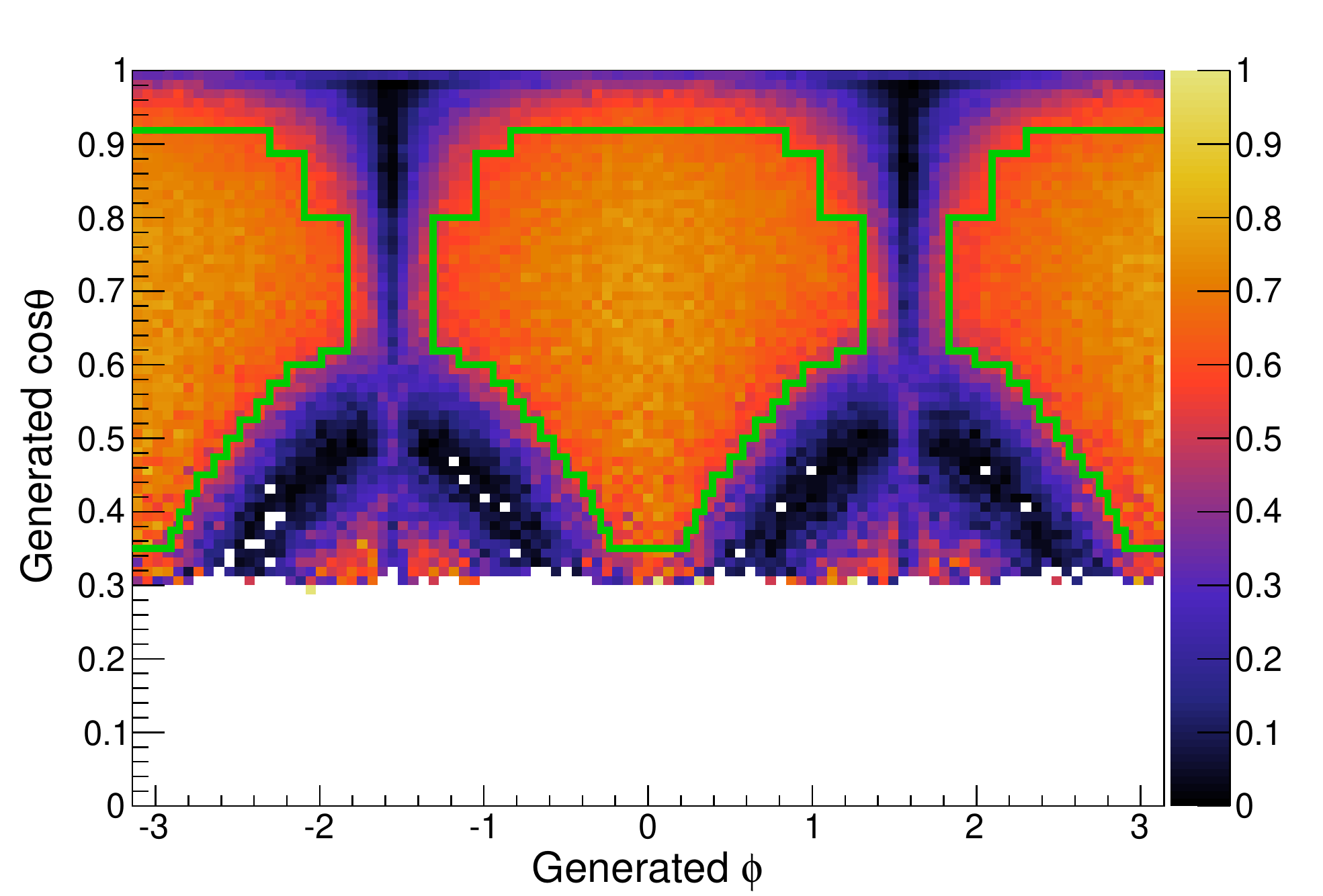}
\includegraphics[width=0.49\textwidth]{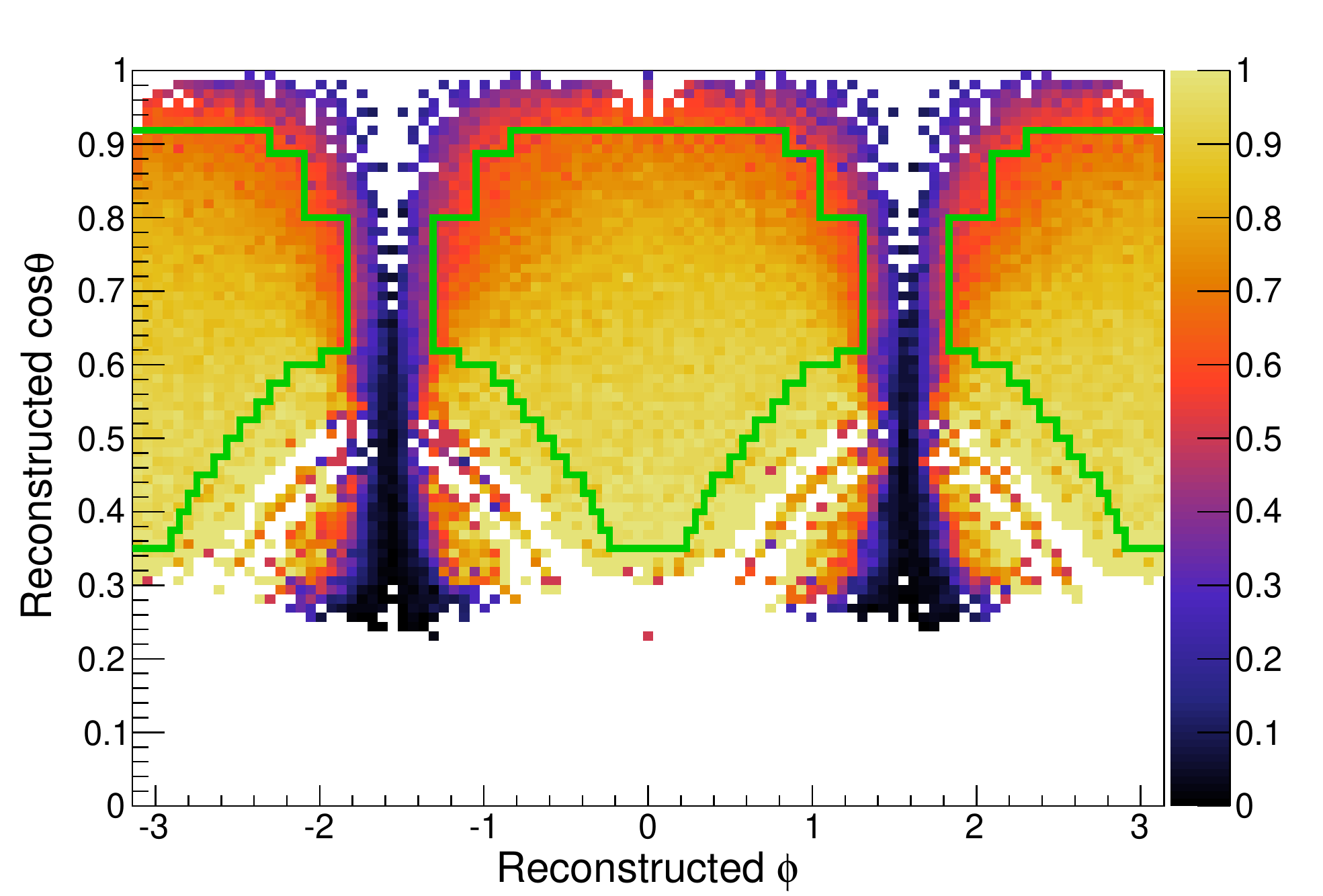}
\caption{Monte Carlo determination of muon track reconstruction accuracy.  The color scale
indicates the probability for the angle to be reconstructed within $5^{\circ}$ of the generated (true)
angle.  This is shown as a function of the generated (left) and reconstructed (right) azimuthal and 
polar angles ($\phi$ and $\theta$, respectively) with respect to vertical.
The angular region of interest (ROI), where reconstruction is deemed sufficiently
accurate, is enclosed by green lines.} \label{fig:muflux_5deg_accuracy}
\end{center}\end{figure}

Vertical muons which pass parallel to the wireplanes do not typically have their direction 
reconstructed accurately, as there is no time difference between the arrival of charge
on the different V- and U-wires.  Poorly reconstructed muons often have their best fit direction
nearly collinear with the U- or V-wires.  To avoid these pathologies, we selected an angular
region of interest (ROI) in which the muon track fitter gives accurate reconstructed angles, and we only used
muons in that region for the flux measurement.  Geant4-based MC studies were used to identify this
well-behaved region, and their results are shown in figure~\ref{fig:muflux_5deg_accuracy}.

\begin{figure}\begin{center}
\includegraphics[width=0.49\textwidth]{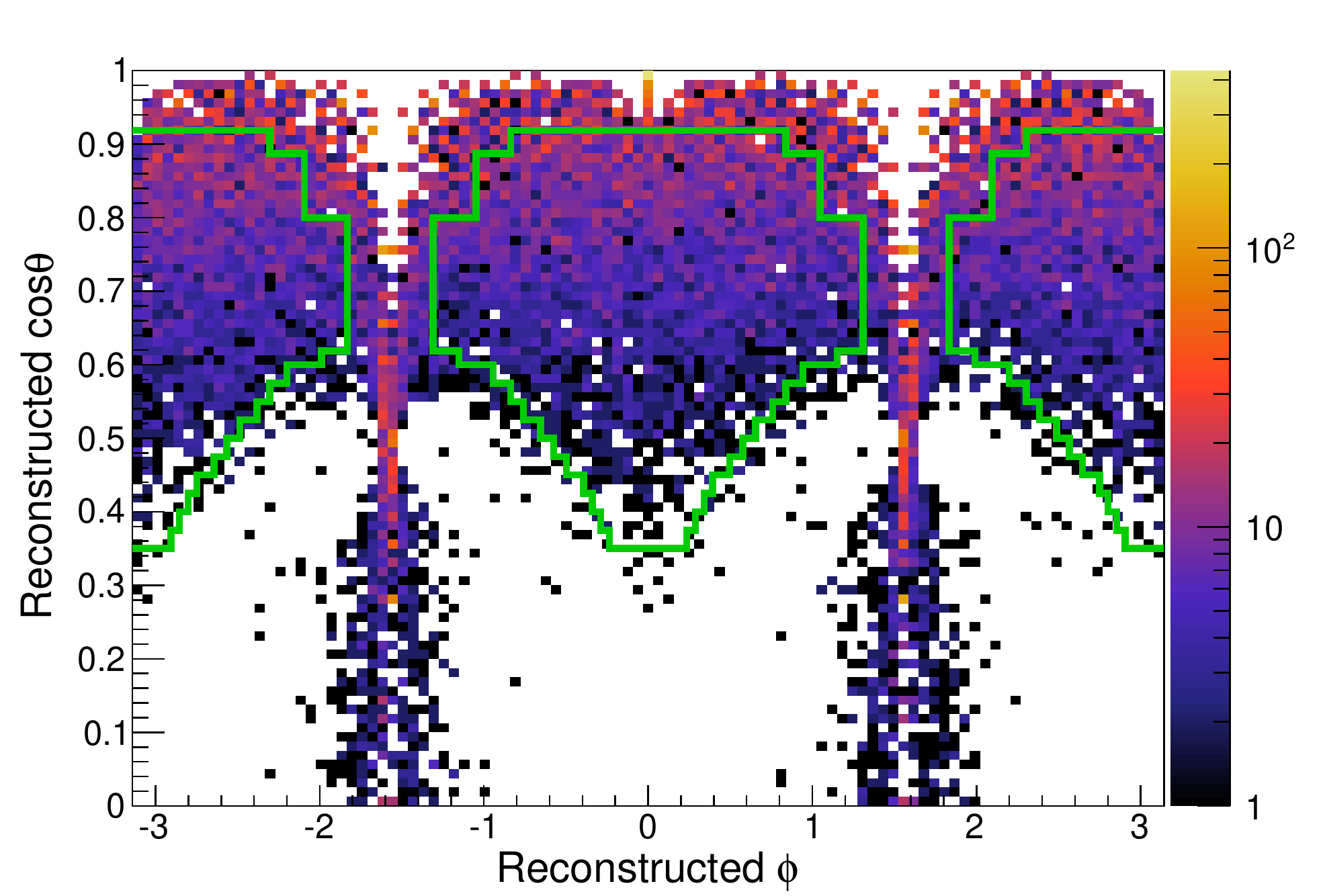}
\includegraphics[width=0.49\textwidth]{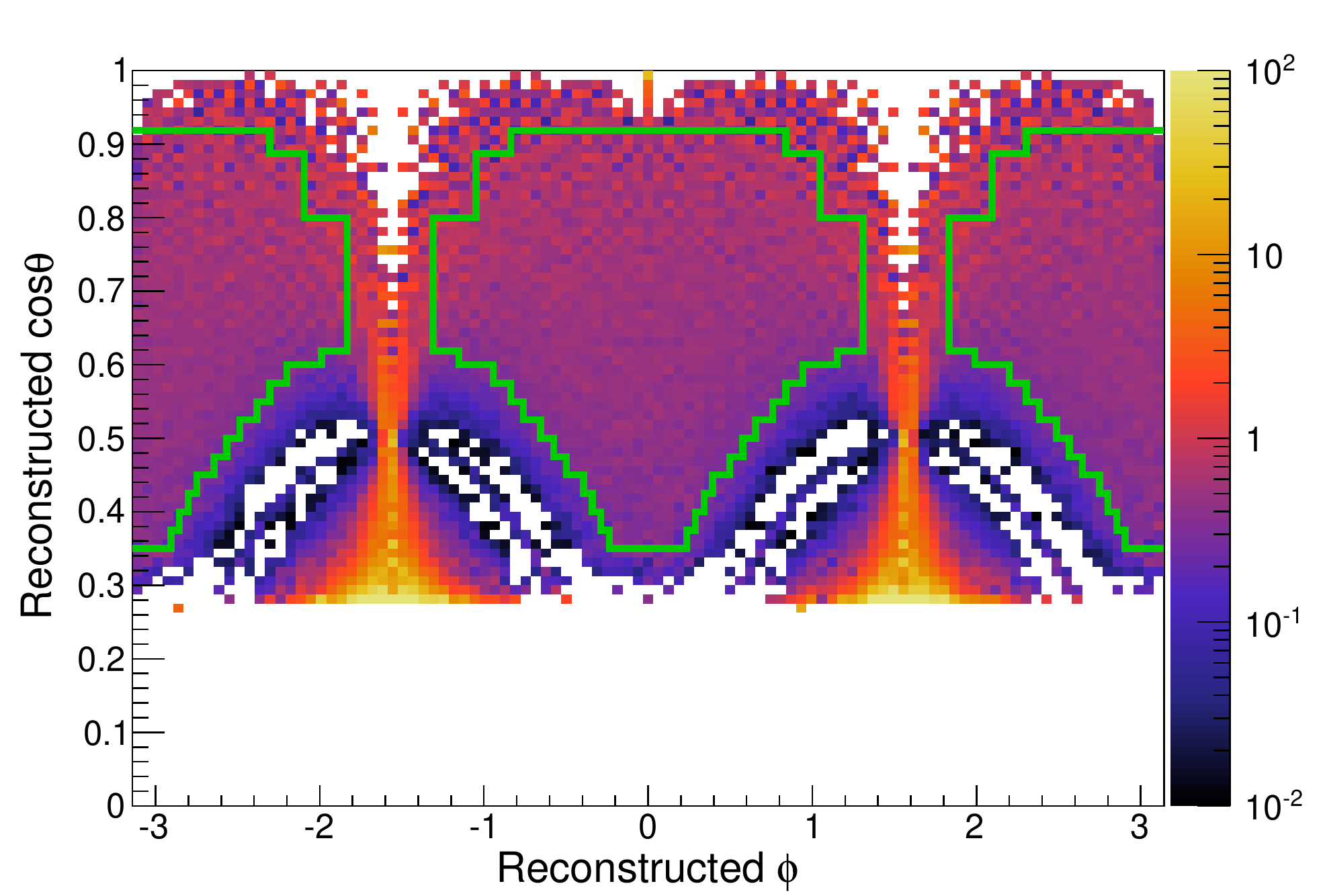}
\caption{
(left) The reconstructed muon directions from data. The color scale represents the 
absolute number of observed counts.  (right) MC events, the color scale represents the ratio of
reconstructed muons in each angular bin to muons simulated with that angle and 
directed at the active LXe.  
The angular region of interest (ROI) is enclosed by green lines.  The excess of
reconstructed muon angles near $\phi = \pm \sfrac{\pi}{2}$ is due to the tendency for mis-reconstructed
muons to be assigned angles parallel to the wireplanes.
}
\label{fig:muflux_data_and_ratio}
\end{center}\end{figure}

Figure~\ref{fig:muflux_data_and_ratio} shows measured muon track directions
and the ratio of
simulated to reconstructed angles from Monte Carlo simulation (MC).  A total of 14810 muons were reconstructed
in the angular ROI, representing 503.118 days livetime.  
The fraction of reconstructed 
muons with their reconstructed angle within the ROI (green-outlined region in 
the left panel of figure~\ref{fig:muflux_data_and_ratio}) is
$43.8\pm0.1\%$ ($43.3\pm0.3\%$) in MC~(data).

\subsection{\label{sec:muon_measurement} Muon flux measurement}

The total flux is calculated as
\begin{equation}
\Phi = \frac{N_{\mathrm{det}}}{\epsilon \Delta T A},
\end{equation}
with $\epsilon$ the efficiency for a muon passing through the generation plane of area $A$
to be detected within the angular ROI, and
$N_{\mathrm{det}}$ the number of such muons detected in time $\Delta T$.
MC studies showed that the fraction of 
muons passing through the generation
plane  (including muons not simulated because of targeting) which are reconstructed in the 
angular ROI is  $ \epsilon = 2.46\pm0.05~\mathrm{(stat)}\times10^{-4}$, before systematic
effects are accounted for.

A summary of our systematic uncertainty estimates and corrections is found in table~\ref{tbl:mufluxsys}.
All systematic uncertainties
are treated as independent, added in quadrature, and applied symmetrically to the result
except for the uncertainty in angular distribution.

\begin{table}[htbp]
\begin{center}
\begin{tabular}{|c|c|c|}
\hline
Systematic & Correction value & Uncertainty value\\
\hline
TPC noise tagging efficiency & $+3.0\%$ & 3.0\%\\
Targeting efficiency & $-0.51\%$ & 0.5\%\\
Finite generation plane correction & $-0.57\%$ & --- \\
Reconstruction efficiency & --- & 0.4\% \\
ROI fraction & --- & 1.1\% \\
MC statistical error & --- &  0.1\% \\
Muon angular distribution & --- & $^{+3.2\%}_{-2.9\%}$ ($I_{\mathrm{v}}$), $^{+1.2\%}_{-0.7\%}$ ($\Phi$)\\
Muon energy distribution & --- & 0.25\% \\
\hline
Total systematic & $+1.89\%$ & $^{+4.6\%}_{-4.4\%}$ ($I_{\mathrm{v}}$), $^{+3.5\%}_{-3.3\%}$ ($\Phi$) \\
\hline
\end{tabular}
\caption{\label{tbl:mufluxsys} Summary of systematic corrections and error estimates.  
All uncertainties are added in quadrature.
Both corrections and uncertainties are listed here.  The vertical 
intensity ($I_{\mathrm{v}}$, see equation~\ref{eqn:vi})
and flux ($\Phi$) calculations are affected by angular distribution uncertainties
differently.
The data statistical uncertainty is 0.8\%.}
\end{center}
\end{table}

The EXO-200 event reconstruction software identifies electronics noise-induced events
with noise-finding algorithms.  
Unphysical signals such as large initial negative signals on the collection wires, 
simultaneous saturation of most channels, and unusual negative pulses in the APDs can 
trigger the noise finder \cite{Herrin:2013iq:appxA}.  Signals appearing to be muon tracks
present in both TPC halves in a way which cannot be explained by a muon crossing the cathode
are also tagged as noise.
In rare cases, legitimate muon signals may be incorrectly tagged as noise, 
leading to a loss in detection efficiency. Our studies showed that the 
noise tagging rate differs slightly for muons in data and MC.  A
correction was obtained by examining data events rejected
as electronics noise and identifying by visual inspection those which appear to be good muon events.
The noise tag cut efficiency for good muons in data (MC) is 96.3\% (99.3\%).
The MC efficiency is corrected to match the data. Although corrected for, the 3\% efficiency difference is taken
as an uncertainty.  We attribute this discrepancy to imperfect
simulation of electronics behavior in MC.

Muons may scatter and therefore not travel on straight lines as assumed by our MC targeting.  
Performing a fit to the number of muons in the ROI as a function of closest distance
of approach to active LXe, we estimated that a 0.51\% correction to the targeting efficiency 
is necessary to account for
contributions from scattered muons.  Conservatively, we also treat this as an additional uncertainty.

The muon efficiency uncertainty was estimated using muons passing through the 
cathode (and thus producing a track in both TPC halves).  The fraction 
of ROI muons in which the muon produces 
 a valid track in the ROI in both TPC halves ($f_{2\times\mathrm{ROI}}$)
is a function of the muon angular distribution and the reconstruction efficiency.  
As the
angular distribution was already constrained by our data, we used the difference in
$f_{2\times\mathrm{ROI}}$ values observed when comparing data and MC to estimate this uncertainty.
The difference in the ratio of ROI muons vs.~total reconstructed muons when comparing
data and MC was used as an additional ROI rate uncertainty.  This likely amounts to
double-counting a single uncertainty, but we treated these as independent, adding them in
quadrature, to be conservative.

Simulations were conducted to test whether any muon secondaries, created in showers, would reach the
TPC and affect the reconstruction efficiency.  No statistically significant effect was observed.
A literature 
review~\cite{Gaisser:1985yw, Cannon:1971jh} showed that
muon bundles are infrequent and diffuse enough to provide negligible impact to our
measurement.

\begin{figure}\begin{center}
\includegraphics[width=0.85\textwidth]{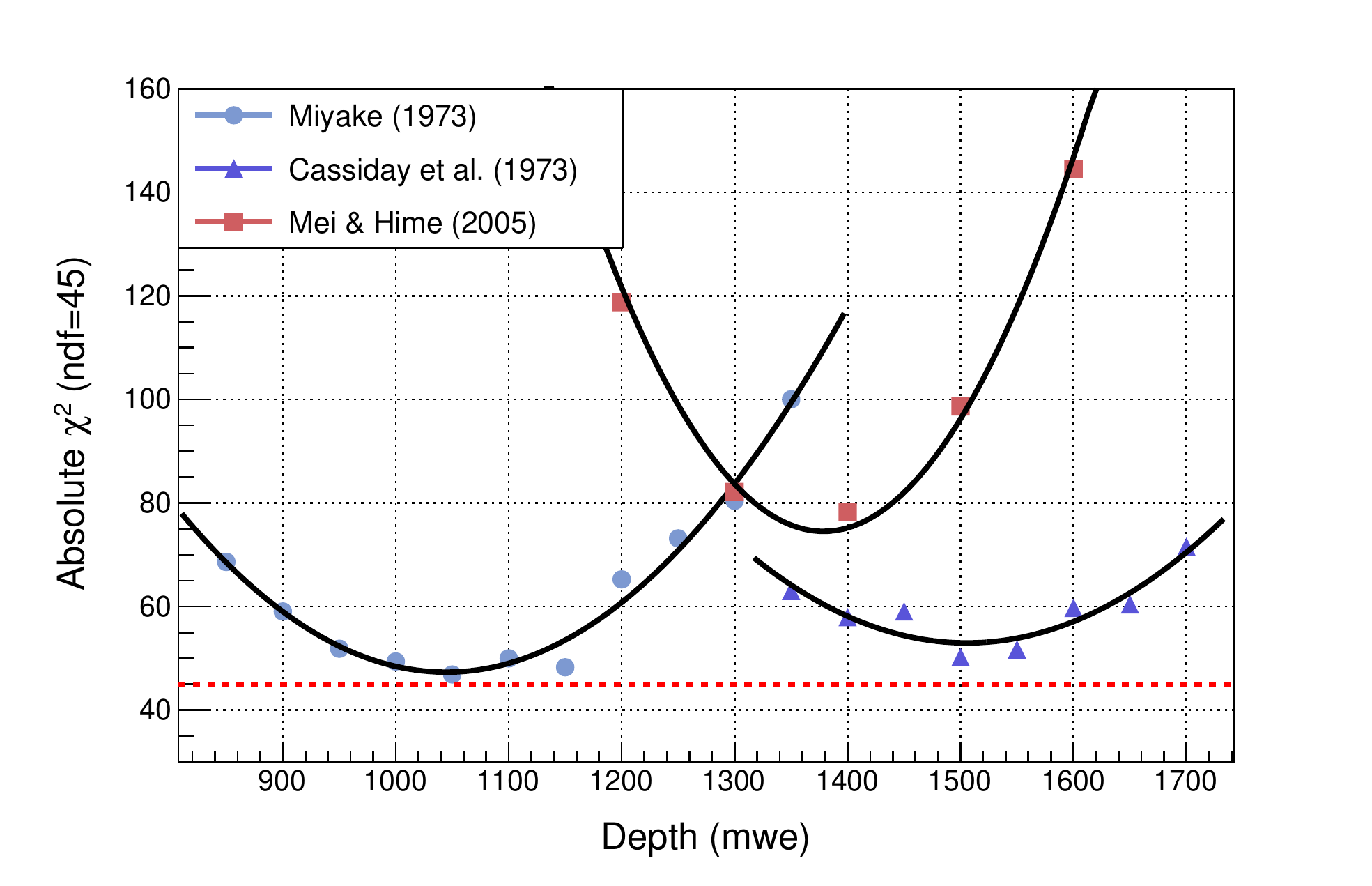}
\caption{
Pearson $\chi^{2}$ between data and MC distributions of reconstructed $\cos\theta$ for ROI muon events.  The
best-fit distribution was Miyake~1050~mwe, although Cassiday et al.~provides a similar 
best $\chi^{2}$ at 1500~mwe.
Distributions from 
Miyake~\cite{Miyake:1973qk}, Mei~\cite{Mei:2005gm}, and Cassiday~\cite{Cassiday:1973tx}
were considered.  Parabolic fits were made to confirm the expected behavior and to better
distinguish the true minima from artifacts due to statistical fluctuations.
}
\label{fig:chi2plot}
\end{center}\end{figure}

The choice of muon angular distribution 
was evaluated by comparing the measured $\cos\theta$ distribution
for events in our angular ROI to MC-produced events using several empirical angular 
distributions (Miyake~\cite{Miyake:1973qk}, Mei~\cite{Mei:2005gm}, and 
Cassiday~\cite{Cassiday:1973tx}).  The phenomenological distributions were
simulated for various values of the depth parameter $h$, at intervals of
50~mwe.  We used a Pearson $\chi^{2}$ statistic to quantify the difference between the measured
and simulated angular distributions.  The results are shown in figure~\ref{fig:chi2plot}.  

\begin{figure}[htpb]
\begin{minipage}{0.48\linewidth}
	\begin{center}
	\includegraphics[keepaspectratio=true,width=3.3in, trim = 0mm 0mm 0mm 0mm]
	{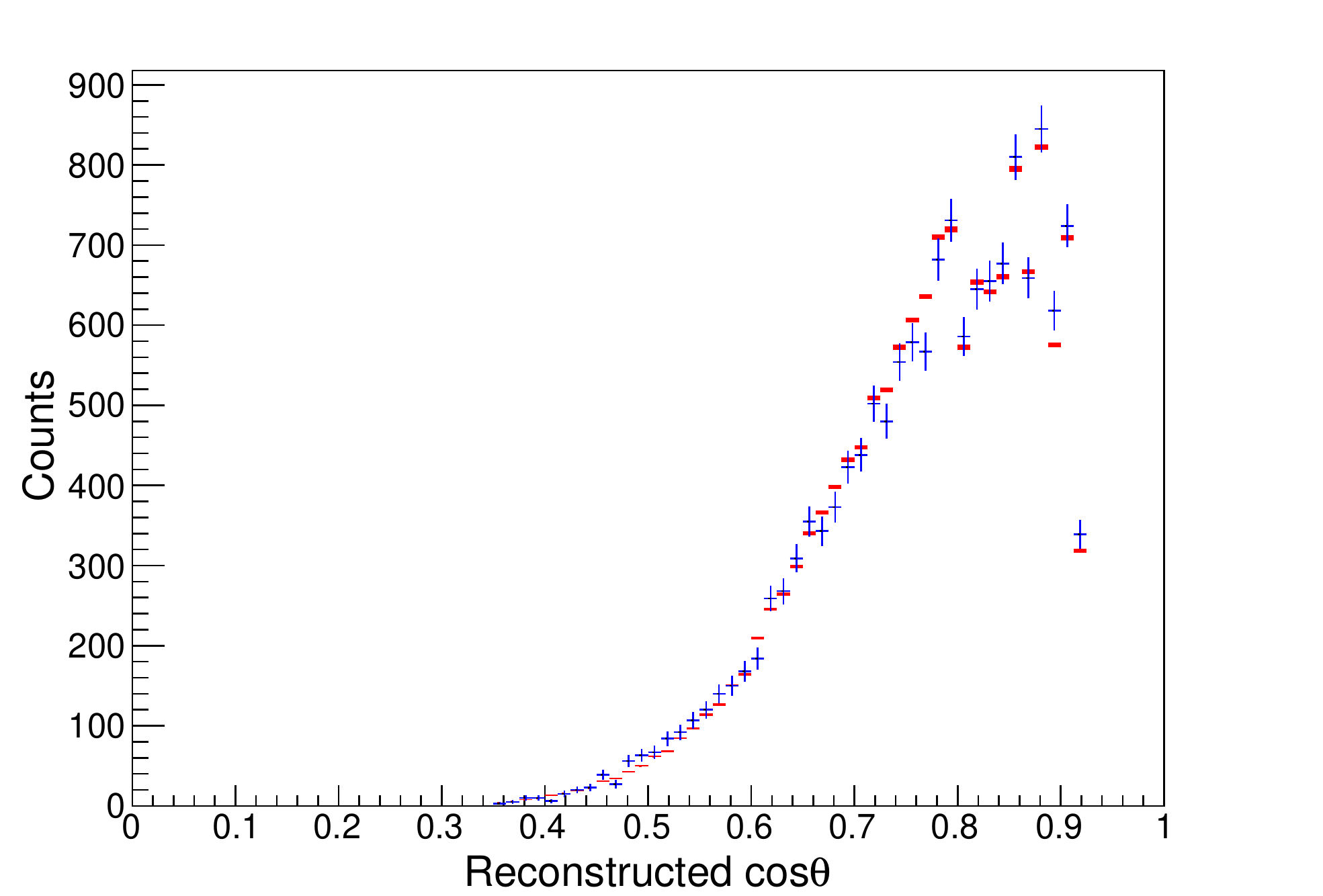}
	\end{center}
\end{minipage}
\begin{minipage}{0.48\linewidth}
	\begin{center}
	\includegraphics[keepaspectratio=true,width=3.3in, trim = 0mm 0mm 0mm 0mm]
	{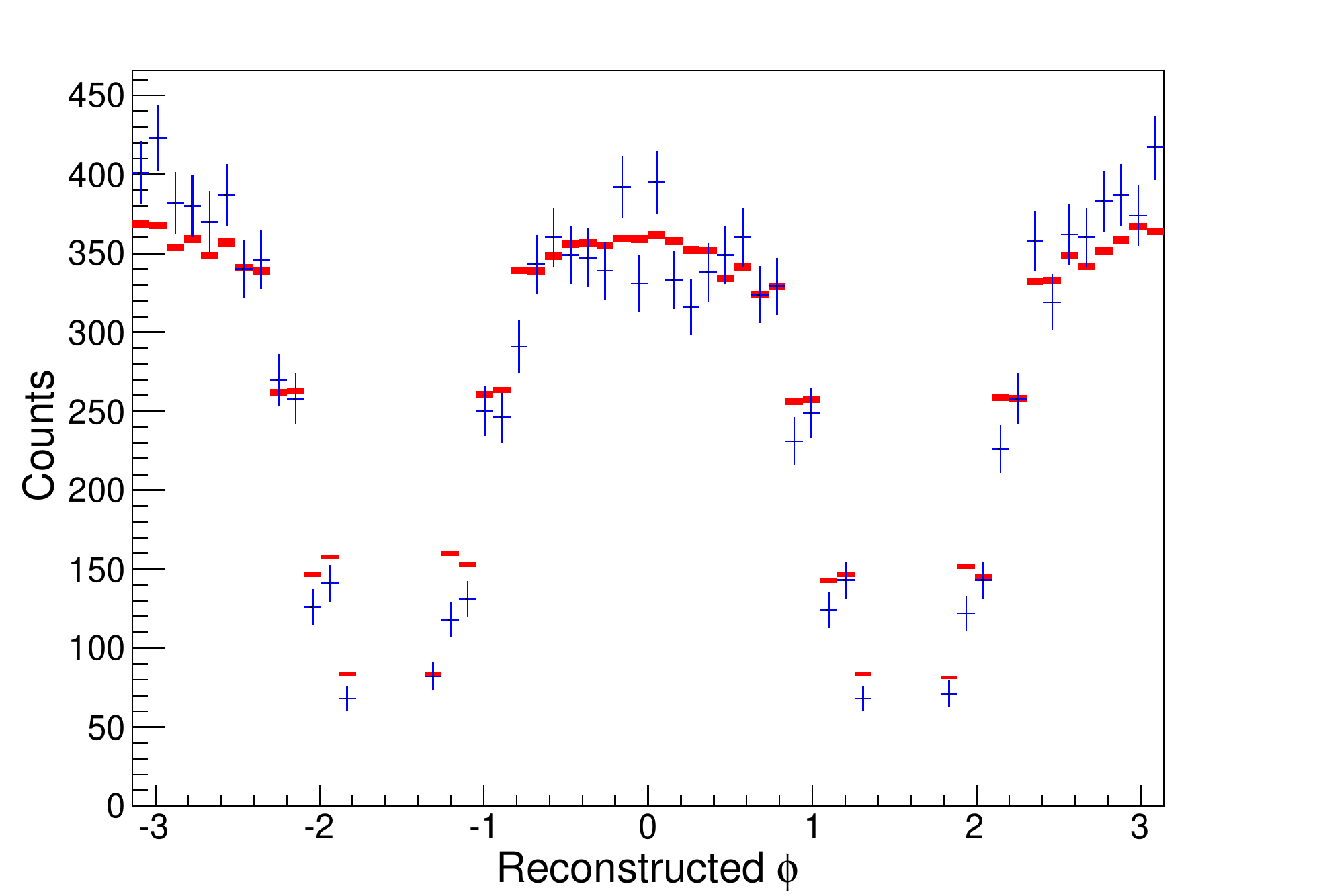}
	\end{center}
\end{minipage}
\caption[Data vs. MC for muon angular distribution]
{Distributions of $\cos\theta$ (left) and $\phi$ (right) for reconstructed muons in the angular ROI.  
Red rectangles are MC described by equation~\ref{eqn:miyake_ang}, blue 
crosses are data.  The distributions are normalized
to correspond to the same number of events in data and MC.}
\label{fig:miyake_e1600_a1050_theta_and_phi}
\end{figure}

The Miyake~\cite{Miyake:1973qk} distribution (equation~\ref{eqn:miyake_ang}) with 
$h=1050$~mwe gave the best agreement.  Reconstructed $\cos\theta$ and $\phi$ distributions 
based on this optimal parameterization are shown overlaid with data in 
figure~\ref{fig:miyake_e1600_a1050_theta_and_phi}.
The Cassiday~\cite{Cassiday:1973tx} distribution gave slightly worse ($\Delta\chi^{2}=3.4$) agreement at a depth parameter
of $h=1500$~mwe, which is closer to the physical depth of WIPP.  As discussed before,
we treat $h$ as a free parameter of a phenomenological distribution, utilized to achieve
good agreement between data and MC.  Because we are directly measuring the angular distribution
and not relying on a parameterization to be accurate,
we can select the best-fitting distribution without concern for the intended physical meaning
of $h$.   We chose to absorb this choice into a systematic
uncertainty due to angular distribution.

If there is an angular bias in reconstruction efficiency that is not present in MC,
the angular distribution will be measured incorrectly.  To account for this possibility, 
the same angular distribution
variation was repeated after skewing the reconstruction efficiencies
in such a way as to reflect the observed $f_{2\times\mathrm{ROI}}$ $\cos\theta$ dependence. 
The resulting best-fit depths for Miyake and Cassiday were used to define a further systematic
uncertainty to the angular distribution.
The central value for flux was calculated assuming the Miyake 1050~mwe distribution with no efficiency skewing.

We also performed a test of the muon energy spectrum underlying the MC model by
varying the $\epsilon_{\mu}$ and $b$ parameters
in equation~\ref{eqn:e0sub} in the ranges of
$500-693$~GeV and $(3.0-4.0)\times10^{-4}~\mathrm{(mwe)^{-1}}$, respectively.
At the extremes, the simulations showed a 0.25\% variation in measured flux, 
which was treated as an additional systematic uncertainty.

Adding all uncertainties in quadrature and applying the described corrections, 
the measured muon flux (flux crossing a flat horizontal surface from above) is

\begin{equation}
\Phi = 4.07\pm0.14~\mathrm{(sys)}\pm0.03~\mathrm{(stat)}\times 10^{-7} \mathrm{cm^{-2}~s^{-1}},
\end{equation}
yielding a vertical muon intensity of 
\begin{equation}
I_{\mathrm{v}} = I_{\mu}\left(\theta=0\right) = 2.97^{+0.14}_{-0.13}~\mathrm{(sys)}\pm0.02~\mathrm{(stat)}\times 10^{-7} \mathrm{cm^{-2}~s^{-1}~sr^{-1}}.
\label{eqn:vi}
\end{equation}

 This corresponds to a standard depth of $1624^{+22}_{-21}$~mwe by 
 the evaluation of \cite{Crouch:1987ICRC}.  These results differ from those published 
 earlier by Esch et al.~\cite{Esch:2004zj}. The new flux measurement presented here is based on 
 a measured angular distribution while the previous measurement assumed 
 the Miyake distribution with a depth parameter now known not to reproduce the data well.

\section{\label{sec:nu_sim}Simulations of cosmogenic nuclide production}

\subsection{\label{sec:mc_generation}MC generation}

Two MC simulation packages, FLUKA~\cite{Bohlen2014211, Ferrari:2005zk} and Geant4~\cite{Allison:2006ve, Agostinelli:2002hh},
were used to model cosmogenic nuclide production.
This approach allows us to 
evaluate the model dependence of neutron production and transport.

The EXO-200 analysis relies on simulations to model the detector response for neutron
capture and all other processes measured.
The estimation of capture $\gamma$ detection efficiencies and calculation of 
probability density functions (PDFs) was performed with the same
EXOsim software as for the muon flux estimation~\cite{Albert:2013gpz}.  
The model has been tuned for accurate low energy electromagnetic physics and incorporates 
particle interaction and electronic signals simulation. 
Different Geant4 physics models were used
for simulating neutron production
and transport, and for estimating cosmogenic nuclide production rates.
In all muon-induced neutron production simulations, the best-fit muon angular 
distribution (Miyake 1050~mwe) was used, 
and the muon energy spectrum was based on the best-fit depth (1624~mwe).

\subsubsection{\label{sec:fluka_sim}FLUKA}

FLUKA version 2011.2c was used for the studies reported here.  The simulation 
geometry is a simplified version of the
geometry used in EXOsim, focused primarily on the general shapes, and geared towards
the proper reproduction of objects of
significant mass (see table~\ref{tab:componentmasses}).  Thus, features such as the cathode and TPC leg supports are neglected.  
The total mass of lead, cryostat, HFE-7000, xenon, and TPC copper are all
simulated to better than 1\% agreement with the Geant4 geometry and CAD reference designs.

The simulated geometry for the WIPP cavern includes
5~m of salt above EXO-200, 3~m
on the floor and near walls, and 2~m thick walls at the ends of a $\pm9$~m long cavern.  
The salt was simulated as pure
NaCl with 1\% H$_{2}$O by molecular fraction \cite{Kuhlman_Malama_2013}.  
Studies of muon shower production in 
FLUKA with this material composition confirmed that 5~m overburden is sufficient for the muon shower to reach
steady state for neutron production, and that neutrons rarely stray more than 3~m
from the muon in a perpendicular direction.  The muons were generated in a 15~m radius plane
above the salt, requiring a correction of $+4.9\%$ to muon-induced activities
to account for the flux expected to originate from beyond the finite generation plane, not
simulated in this model.

The FLUKA simulation used the PRECISIO(n) defaults, with several other options activated
for accurate treatment of cosmogenic secondaries.  To achieve a detailed modeling of
nuclear de-excitation we utilized EVAPORAT(ion) and COALESCE(nce).  We activated 
photonuclear interactions
and muon-muon pair production with PHOTONUC, heavy ion transport with IONTRANS, 
electromagnetic dissociation of heavy ions with EM-DISSO(ciation), and the 
PEANUT hadronic generator model at all energies with PEATHRES(hold).  The 
rQMD-2.4~\cite{Sorge:1989dy} and
DPMJET-3.0~\cite{Roesler:2000he} models were activated.  We additionally
set electromagnetic transport cutoffs of 2~MeV and 600~keV for $e^{\pm}$ and $\gamma$, 
respectively.  All materials were associated with low energy neutron cross-sections assuming 
they have a temperature of 296~K.  
A study was performed comparing capture rates with
87~K cross-sections and 296~K cross-sections for materials within the cryostat, 
and a correction to approximate 167~K (the temperature of LXe) was derived by 
interpolating the neutron capture rates between 
these temperatures.

\subsubsection{\label{sec:geant4_sim}Geant4}

For the neutron-related studies, Geant4~\cite{Allison:2006ve, Agostinelli:2002hh}
version 10.0.p02 was used. Photo-nuclear, electro-nuclear, muon-nuclear 
interactions and capture of muons on nuclei were enabled. 
The ``QGSP\!\textunderscore BERT'' model was used for hadronic interactions. A package of 
high-precision
models for neutron interactions below 20~MeV was enabled and 
neutron data files with thermal cross sections of version 4.4 were used.  An unreleased
patch~\cite{G4bugzilla} was applied to correct photo-nuclear interaction rates.

Muons were generated with the same energy and angular distribution as was used with
the FLUKA simulation.  In this simulation, no generation plane was utilized.  Rather, all
muons passing through a sphere of 2~m radius, centered on the TPC, were simulated.  The muons
were generated from a spherical surface with 8~m radius, also centered on the TPC.

The masses for each of the major components in both simulations are listed in table~\ref{tab:componentmasses}.

\begin{table}[htbp]
\begin{center}
\begin{tabular}{|c|d{5.2}|d{5.2}|d{5.2}|}\hline
    Material  & \multicolumn{1}{C{1.7cm}|}{Geant4 \newline mass (kg)}  &  \multicolumn{1}{C{1.7cm}|}{FLUKA \newline mass (kg)}  & \multicolumn{1}{C{1.8cm}|}{Reference \newline mass (kg)}  \\ \hline
    TPC LXe  & 169.2  & 168.7 & 168.6 \\
    TPC vessel Cu    &  22.3  &  22.3 &  22.4  \\
    TPC internal Cu & 10.2 &  10.3 & 10.4  \\
    HFE-7000 & 4138  &  4138  & 4140 \\
    Inner Cryo & 2881  &  2841 & 2843  \\
    Outer Cryo  & 3480  &  3451 & 3453  \\
    Main Pb shielding & 55060 &  54920 & 55000   \\
    Outer Pb wall & 8620  &  8629 & 8634   \\
\hline
\end{tabular}
\caption[FLUKA masses vs. reference and EXOSim]{\label{tab:componentmasses} 
Major component masses from Geant4, FLUKA, 
and engineering-model reference values~\cite{Albert:2013gpz, Auger:2012gs}.
Mass agreement between the Geant4, FLUKA, and CAD (reference) geometries is at the 
$\pm1\%$ level.
TPC LXe includes all liquid xenon in the TPC vessel, including xenon outside the teflon
reflector or behind the collection wires.  Xenon in the TPC support legs
or elsewhere in the circulation loop is not included in the mass accounting.
TPC internal Cu includes all copper
parts within the TPC vessel.
}
\end{center}
\end{table}

\subsection{\label{sec:residuals}Muon-induced activation}

Many unstable nuclides are produced directly or indirectly by cosmogenic muons in the detector and shielding
materials. Most of these nuclides are irrelevant because, given the attenuated muon flux underground,
they are produced at very low rates.
We considered unstable nuclides capable of producing events 
with energy near the \znbb Q-value. 
The rate of such events is effectively a product of muon flux, the probability to create 
specific nuclei per muon,
and the efficiency of the subsequent decay to deposit energy in the \znbb energy ROI.

A list of nuclides of interest
is shown in table~\ref{tbl:isotopes}.
Most of these nuclides are produced in xenon, as $\beta$ decays in xenon are especially
important sources of background for \znbb searches.  $^{135}\mathrm{Xe}$ is included in this table
due to its high production rate, although it cannot mimic \otsx \znbb.   
$^{60}\mathrm{Co}$ contributes to \znbb background via coincidence summation of its 
two principal $\gamma$s.  Therefore, $^{60}\mathrm{Co}$ is only a problem in detector 
components in and near the TPC.  The underground, steady-state production of 
$^{60}\mathrm{Co}$ in copper provides a negligible
contribution compared to the left-over activity from above-ground exposure~\cite{Albert:2015nta}.
The total number of atoms of each unstable nuclide generated per unit time is noted, along with
two efficiencies.  We have evaluated the efficiency to deposit between 2.3 and 2.6~MeV in the
xenon fiducial volume (FV) and, in addition, the efficiency for an event to pass 
the EXO-200 \znbb selection cuts, which require SS event topology and an energy deposition in the 
$Q_{\beta\beta}\pm 2\sigma$ energy window (the \znbb ROI)~\cite{Albert:2014awa}.
All cosmogenic radionuclides other than $^{137}\mathrm{Xe}$ make a negligible contribution
to the SS \znbb ROI backgrounds.
Each is calculated to contribute less than 1\% of the contribution from $^{137}\mathrm{Xe}$.
An expanded table of nuclides of interest is found in appendix~\ref{sec:expanded_table}.

\begin{table}[htbp]
\centering
\begin{tabular}{|c c c c c c c|}
\hline
  \multicolumn{1}{|C{2.5cm}}{Region/ \newline Nuclide} & \multicolumn{1}{C{1.1cm}}{Half-\newline life} 
   & \multicolumn{1}{C{1.75cm}}{Total energy (keV)} & \multicolumn{1}{C{1.4cm}}{2.3--2.6 MeV \newline efficiency}& \multicolumn{1}{C{1.7cm}}{EXO-200 SS ROI efficiency} 
   & \multicolumn{1}{C{1.60cm}}{Geant4 rate (yr$^{-1}$)} & \multicolumn{1}{C{1.45cm}|}{FLUKA rate (yr$^{-1}$)} \\

\hline 
HFE-7000 {}\(^{16}\)N   & 7.1~s   & 10420 & 0.015\% & 0.001\% & $2380\pm90$ & $2910\pm110$ \\
\hline
TPC vessel {}\(^{60}\)Co  & 5.3~y & 2823  & 0.19\% & 0.0002\% & $2.6\pm0.3$ & $2.9\pm0.6$  \\
\hline 
TPC LXe {}\(^{130}\)I  & 12~h  & 2949  & 8.5\%  & 0.001\%  & $7.3\pm0.5$    & $21.6\pm1.8$   \\
TPC LXe {}\(^{132}\)I  & 2.3~h & 3581  & 5.5\%  & 0.013\%  & $7.7\pm0.5$    & $22.2\pm1.8$   \\
TPC LXe {}\(^{134}\)I  & 53~m  & 4175  & 4.5\%  & 0.012\%  & $7.3\pm0.5$    & $20.4\pm1.7$   \\
TPC LXe {}\(^{135}\)I  & 6.6~h & 2627  & 2.9\%  & 0.035\%  & $8.6\pm0.6$    & $21.6\pm1.8$   \\
TPC LXe {}\(^{135}\)Xe & 9.1~h & 1165         & ---   & ---    & $1110\pm40$     & $1060\pm40$   \\
TPC LXe {}\(^{137}\)Xe & 3.8~m & 4173         & 4.1\%  & 1.5\%  & $439\pm17$     & $403\pm16$    \\
\hline
\end{tabular}
\caption{\label{tbl:isotopes} Summary of cosmogenic radionuclides of interest.  
All listed nuclides decay through $\beta$-emission, some with
associated $\gamma$s.  The half-lives and total decay energy are 
listed~\cite{Tilley:1993zz, Browne20131849, SINGH200133, Khazov2005497, Sonzogni20041, Singh2008517, Browne20072173}.  
The efficiencies listed have been estimated using EXOsim.
The Geant4 and FLUKA columns
give the estimated production rate for the listed nuclide in atoms per year.  
Iodine efficiencies listed do not
account for iodine removal during recirculation.  Only the
$^{137}\mathrm{Xe}$ production
is dominated by thermal neutron capture, so the FLUKA temperature correction has been applied
only to that nuclide. Uncertainties include the MC statistical uncertainty and the
muon flux uncertainty.
Significant differences in the predicted rates from FLUKA and Geant4 were seen for nuclides 
(such as the iodine isotopes)
which are primarily produced with high energy ($E > 10$~MeV) interactions.
 }
\end{table}

All daughters of the initial radionuclides are either stable or have
no ability to contribute near the \znbb Q-value.  We also considered known metastable
states for nuclides produced in xenon, and found that none of them contribute significantly to backgrounds.

The background contribution from iodine isotopes is almost certainly over-estimated, 
because iodine atoms are removed by the continuous xenon purification.  
It is difficult to calculate the xenon circulation patterns, so 
this effect is not considered here.  Cosmogenic nuclides
produced in the liquid xenon outside the TPC were also studied.  A 
higher rate of activation was found nearer to the outside of the cryostat, but its 
contribution to \znbb backgrounds was found to be negligible, due to the low mass of xenon 
outside the TPC and a flow rate which
allows most $^{137}\mathrm{Xe}$ to decay before reaching the TPC.

\subsection{\label{sec:ncap_gammas}Prompt $\gamma$s following neutron capture}

Veto-tagged data studies (section~\ref{sec:veto_tagged_ana}) require accurate
modeling of signals due to 
prompt $\gamma$s from neutron captures on detector materials.
While FLUKA and Geant4 include models for simulating prompt $\gamma$s after neutron capture,
better accuracy can be achieved using custom capture $\gamma$ cascades.
These cascades were used in EXOsim to generate PDFs for fits to data.
In each case, the $\gamma$s in the cascade are distributed isotropically, and, with the 
exception of HFE-7000, captures were generated uniformly in each volume.
Our simulations showed that thermal neutron capture is the dominant event class in the
data, so the cascades were all based on thermal capture measurements.

\begin{itemize}
\item $^{137}\mathrm{Xe}$ atoms are produced in a capture state
with excited state energy $4025.46\pm0.27$~keV~\cite{mughabghab2006atlas}.  
They relax into the ground state promptly, primarily
through $\gamma$ emission. The main $^{136}\mathrm{Xe}(n,\gamma)^{137}\mathrm{Xe}$ 
cascade was modeled using 
thermal neutron capture data by Prussin et al. \cite{Prussin:1977zz}.  The evaluated
$\gamma$ and level data were converted into a de-excitation chain model to generate the
cascade.  For levels where an imbalance exists
between incoming and outgoing $\gamma$ intensity, it was assumed that an unknown ``continuum''
state exists, and the imbalance was resolved with an additional transition to a randomly
chosen level in the continuum.  To evaluate possible systematic uncertainties in this important capture
cascade, a simulation was conducted based on preliminary data from a measurement of
$^{136}\mathrm{Xe}(n,\gamma)^{137}\mathrm{Xe}$ at the Detector for Advanced Neutron
Capture Experiments.  The DICEBOX code \cite{Becvar1998434} was used to help model the cascade.  
Efficiency differences between this model and the primary model are treated as a systematic
uncertainty.
\item $^{1}\mathrm{H}(n,\gamma)^{2}\mathrm{H}$ results in the emission of a single $\gamma$ with an energy of
2223~keV.  As the HFE-7000 volume is much thicker than the thermal neutron interaction length, we
divided its volume into 6 concentric cylindrical shells of roughly equal thickness 
for simulation in EXOsim and FLUKA, to account for the varying
capture rates and spectral variations at different positions.
\item The capture cascades for $^{63,65}\mathrm{Cu}(n,\gamma)^{64,66}\mathrm{Cu}$ were
based on levels and $\gamma$s extracted from the ENSDF 
database~\cite{Singh2007197, Browne20101093}.  Level transitions
were matched by automated parsing of the data file~\cite{tuliensdfformat}, producing code
for modeling the cascade.
\item The capture cascade for $^{19}\mathrm{F}(n,\gamma)^{20}\mathrm{F}$ was similarly
based on levels and $\gamma$s extracted from the ENSDF database \cite{Tilley1998249}.
\item We found no publications specifically measuring the branching ratios of the 
capture cascade for 
$^{134}\mathrm{Xe}(n,\gamma)^{135}\mathrm{Xe}$, 
so we used the cascade model included in FLUKA.  This cascade is based on a 
study of neutron capture in natural xenon~\cite{fluka134xe}.  Including this PDF
improves the overall fit to veto-tagged data, yet has only a small effect on the final fitted rate
of capture on $^{1}\mathrm{H}$ and \otsx, so cascade model uncertainties do not
significantly affect the results.
\end{itemize}

Our simulations showed that only captures on $^{134,136}\mathrm{Xe}$ (in the TPC), 
$^{63,65}\mathrm{Cu}$ (in the TPC vessel, inner and outer cryostats), and $^{19}\mathrm{F}$
and $^{1}\mathrm{H}$ (in the HFE-7000) contribute significantly to 
veto-tagged neutron capture signals
in EXO-200.  Other possible captures, such as on carbon in the HFE-7000 or on the lead shield,
were neglected, due to a combination of low cross-sections, low masses,
 and/or low $\gamma$ detection efficiency in the TPC.

\section{\label{sec:veto_tagged_ana}Veto-tagged data}

By selecting events collected shortly after any of the 29 muon veto panels trigger, we obtained a 
``neutron-enriched'' dataset.  While not all veto panel triggers are associated with 
neutron events, the narrow time cut rejects the vast majority of 
non-cosmogenic events, leaving little other than neutron capture signals.

\begin{figure}[htb]
	\begin{center}
    \includegraphics[width=0.89\textwidth]{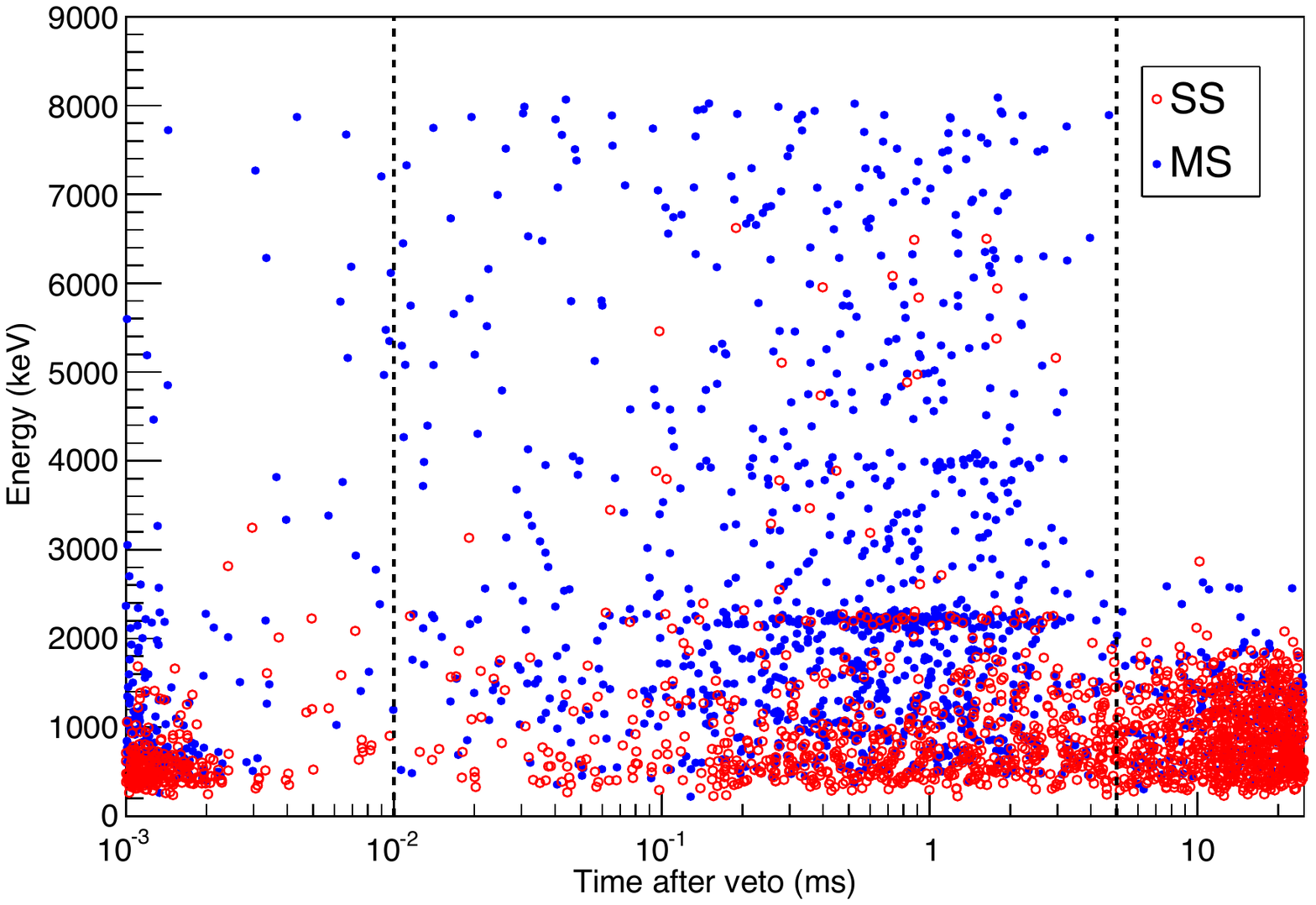}
    \caption[Energy vs. time after veto]
    {Data events occurring shortly after muon veto panel triggers.  The event selection criteria are defined in 
    Section~\ref{sec:data_selection}. Filled (open) points represent MS (SS) events.  
    The time axis is logarithmic.  Signals from capture on $^{1}\mathrm{H}$
    (population at 2223~keV),
    \otsx (population at 4025~keV), and $^{63,65}\mathrm{Cu}$ (spectrum cutting off at 7916~keV) 
    are visible in the data.  The
    dashed vertical lines indicate the selection used to obtain the veto-tagged data set.
    At very short time differences, signals from muon-related processes other than thermal neutron capture dominate.  
    At long time differences, radioactivity-induced events (largely SS
    \tnbb events) dominate.}
    \label{fig:ss_ms_time_overlay_log_dashedline_fixed}
    \end{center}
\end{figure}

Figure~\ref{fig:ss_ms_time_overlay_log_dashedline_fixed} shows the data in and around the veto cut.  
We selected TPC events following a veto-panel trigger by
times between 10~$\mu$s and 5~ms.  This lower bound rejects signals associated directly 
with the muons (such as bremsstrahlung $\gamma$s)
while keeping the vast majority of neutron capture events. There are very few random coincidences with
non-cosmogenic
radioactive decay.  This time window was chosen based on studies of signals from neutron
capture on $^{1}\mathrm{H}$, which produces a 2223~keV $\gamma$.  Other prominent capture signals 
in the tagged data include $^{136}\mathrm{Xe}(n,\gamma)^{137}\mathrm{Xe}$ (sum peak line at 4025~keV) 
and $^{63,65}\mathrm{Cu}(n,\gamma)^{64,66}\mathrm{Cu}$ ($\gamma$ spectrum up to 7916~keV).

\subsection{\label{sec:data_selection}Data reconstruction and selection}
\begin{figure}[htpb]
	\begin{center}
	\includegraphics[keepaspectratio=true,width=4.7in, trim = 0mm 0mm 0mm 0mm]{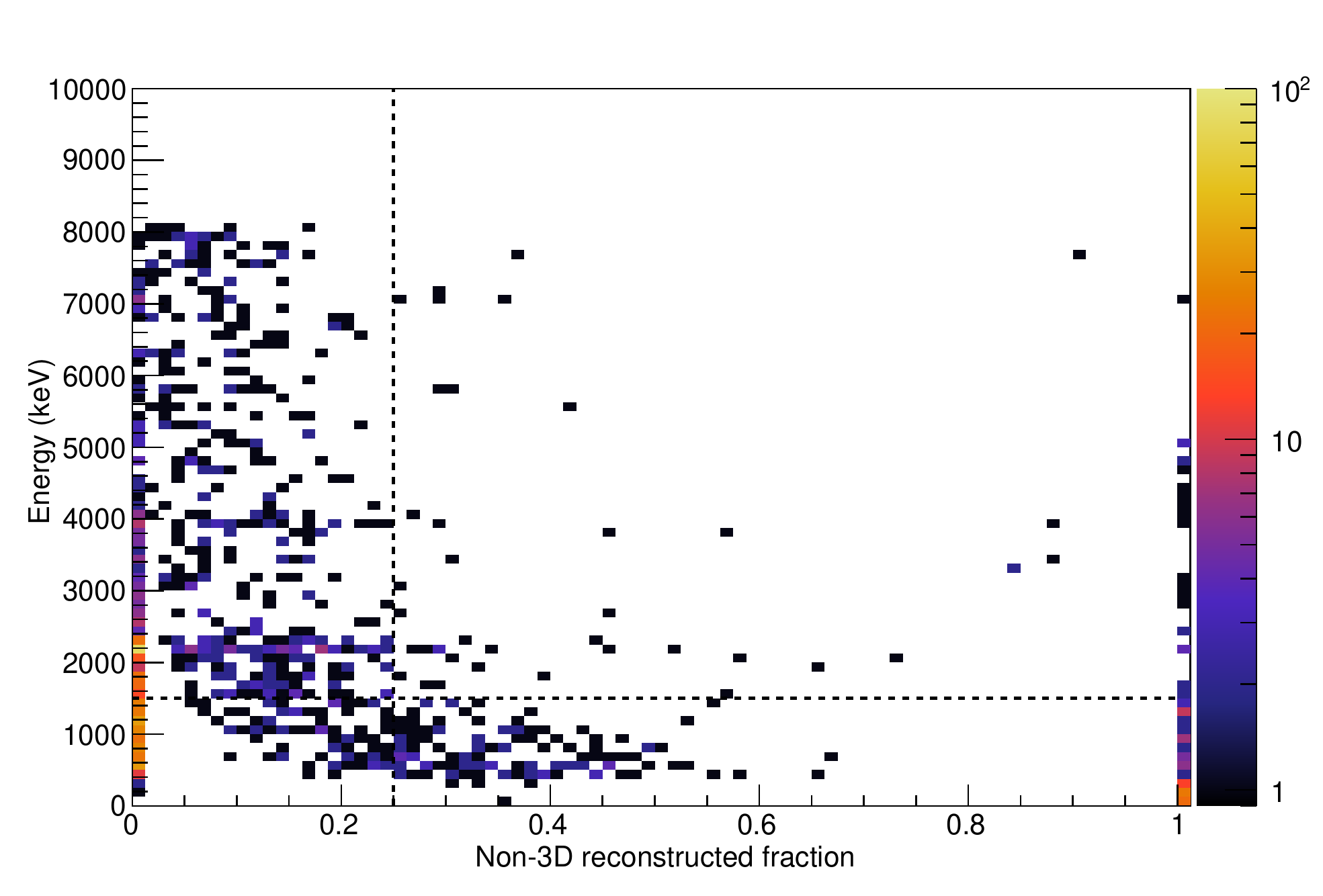}
	\caption[3D reconstruction fraction]{Event energy vs.~non-3D reconstructed fraction
	for veto-tagged data. 
	The color scale indicates the number of MS events in a bin.  
	Events to the right of the vertical dashed line or below the horizontal one 
	were rejected.  The
	far left bins are fully-3D reconstructed events, and the far right
	bins have no charge clusters with a 3D position.}
	\label{fig:3Dreconfrac}
	\end{center}
\end{figure}

This analysis used the data set and analysis techniques of the recent \znbb analysis \cite{Albert:2014awa},
but with modified event reconstruction and selection.
Briefly, the \znbb analysis
uses energy depositions in the TPC fiducial volume from isolated
$\gamma$ and $\beta$ emissions to identify backgrounds
and possible signals.
Changes were focused on improving the efficiency for detecting
capture $\gamma$s (as opposed
to selecting the \znbb signal).  Reconstruction changes include:
\begin{itemize}
\item The signal-finding and clustering algorithms were modified to better identify 
signals sharing frames with other signals, either from capture $\gamma$s or TPC muons.
\item Denoising  \cite{Albert:2014awa} of APD signals was not used.  The denoising algorithm
was developed for sparse events and does not cope well with short times between scintillation events.  
Choosing to forgo
the denoising results in higher acceptance at the cost of slightly degraded energy resolution.
\end{itemize}
Event selection changes include:
\begin{itemize}
\item The fiducial volume was expanded slightly toward the cathode, 
requiring $182~\mathrm{mm} > |z| > 5~\mathrm{mm}$ with a hexagonal cut in $x$-$y$ with a
162~mm apothem.
\item Events containing clusters without a full 3D position were accepted.  In cases where a signal on
a U-wire cannot be matched with that of a V-wire, the $x-y$ position of the charge cluster 
cannot be determined.  Rather than reject such events, we required that at least
75\% of the total charge cluster energy be found in fully 3D reconstructed
clusters.  This greatly boosts efficiency for events with multiple Compton scatters.  
Figure~\ref{fig:3Dreconfrac} shows the 
fully 3D reconstructed fraction for MS events.  
The FV cut was only enforced along the $z$-direction (drift direction) for charge clusters
without a determined $x-y$ position.
\item Events with more than one scintillation cluster ($N_{\mathrm{scint}}>1)$ in the data 
frame were accepted.  Scintillation clusters within 119~$\mu\mathrm{s}$ (a drift time plus uncertainty) of other 
scintillation clusters were rejected, as the charge drift times 
may overlap.
\item No veto rejection was applied for events close in time to TPC muons, muon veto panel triggers, or
other TPC events.
\end{itemize}

\begin{figure}[htpb]
	\begin{center}
	\includegraphics[keepaspectratio=true,width=4.7in, trim = 0mm 0mm 0mm 0mm]{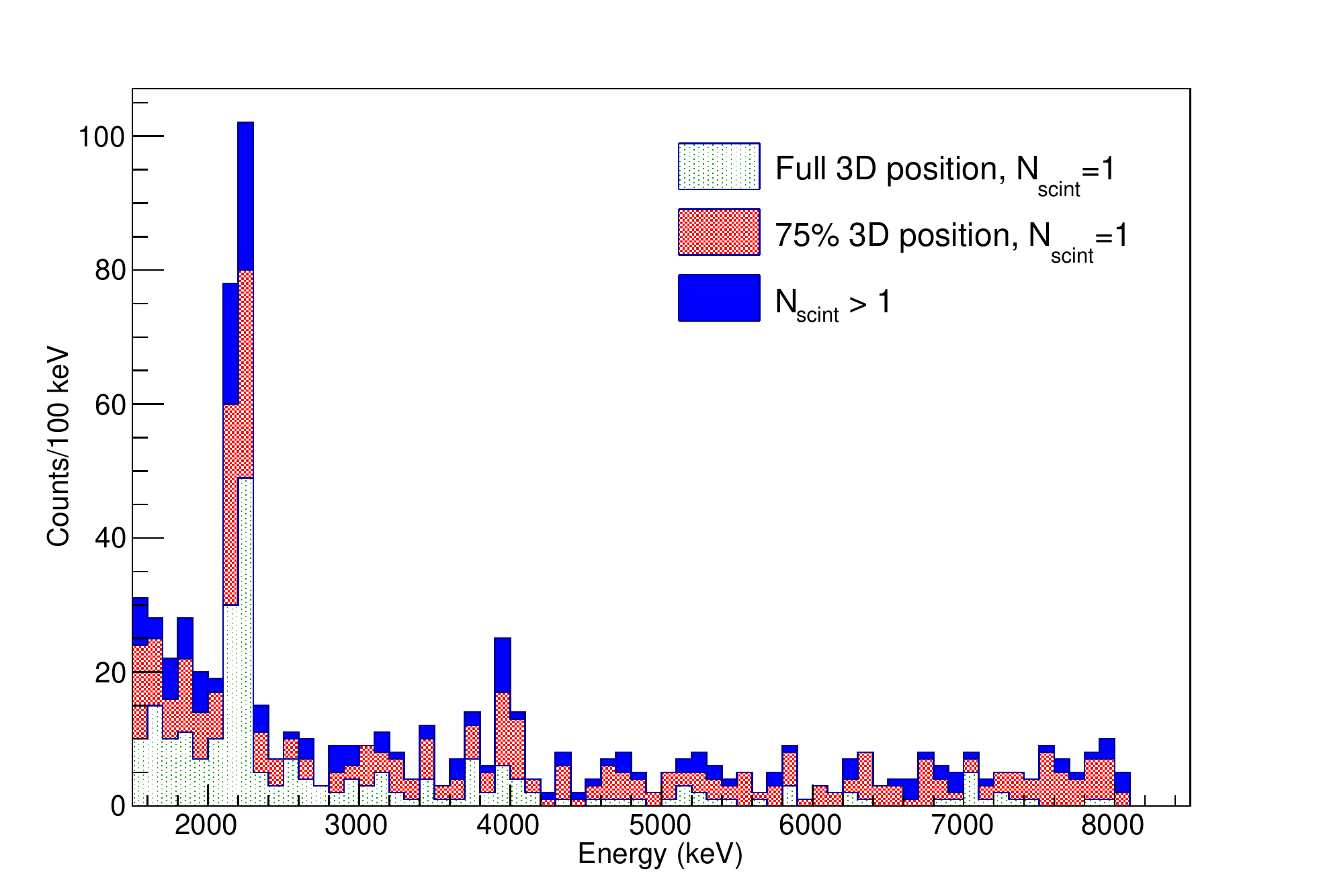}
	\caption[TPC] {Histogram showing the effect of relaxing the 3D reconstruction
	cut and the $N_{\mathrm{scint}}=1$ cut on MS veto-tagged data.}
	\label{fig:newcuts_stack}
	\end{center}
\end{figure}

Figure~\ref{fig:newcuts_stack} shows how the MS dataset grew with the addition of 
multi-scintillation cluster events and the relaxing of the 3D reconstruction cut.  These
modifications more than doubled the efficiency for neutron capture events, relative to
the \znbb event selection criteria.

\subsection{\label{sec:fitting}Fit to data}

To measure the rate of neutron capture on each of the detector components, we performed a
maximum-likelihood fit simultaneously in the SS and MS event energy and in
the standoff distance parameter (nearest
distance between a charge cluster and the teflon reflector or anode plane).  The analysis
threshold was 1500~keV, chosen to minimize complications related to partially 3D 
reconstructed events.  We fitted with
six fully independent PDFs: capture on \otsx, $^{134}\mathrm{Xe}$, $^{1}\mathrm{H}$, 
$^{19}\mathrm{F}$, and $^{63, 65}\mathrm{Cu}$, plus one more PDF accounting for radioactivity-induced
background events.  The components of this background PDF were constrained to 
have relative contributions as measured using a fit
of non-veto-tagged data.  The HFE-7000 capture PDFs were composed of component PDFs representing
six concentric geometric
shells (see section~\ref{sec:ncap_gammas}).  The copper capture PDF 
was also composed of six component PDFs, describing neutron capture on $^{63}\mathrm{Cu}$ and $^{65}\mathrm{Cu}$ in
TPC vessel, inner cryostat, and outer cryostat.  
The fractional contribution of each of the
HFE-7000 shells or copper components to their sum PDFs 
were determined by MC predictions, and were constrained with a 20\% gaussian uncertainty during
fitting.  The constraint was used as these component PDFs are degenerate with each other 
in energy and standoff parameter,
so the fitting process itself has little power in resolving, for example, the relative
capture rates on the TPC vessel and cryostat.

We fitted using the same techniques used in previous 
analyses~\cite{Albert:2014awa, Albert:2013gpz}, including
a SS fraction constraint (constraining the event number ratio
(SS/SS$+$MS) for each PDF), and an overall normalization constraint (constraining
detector efficiency).  These are gaussian constraints with widths of 5.3\% and 10.5\% respectively,
and represent uncertainties derived from source data studies.  
The $\beta$-scale (summarized by the parameter $B$, the correction
needed to convert calibrated $\gamma$ energies into $\beta$ energies,
$E_{\beta} = B E_{\gamma}$~\cite{Albert:2014awa}) was determined 
with a fit to the non-veto-tagged data, yielding $B = 0.998\pm0.004$.  This parameter was
fixed to the central value, though it had little effect as the veto-tagged data 
is dominated by capture $\gamma$s.

\begin{figure}[htpb]
	\begin{center}
	\includegraphics[keepaspectratio=true,width=6.2in, trim = 0mm 0mm 0mm 0mm]{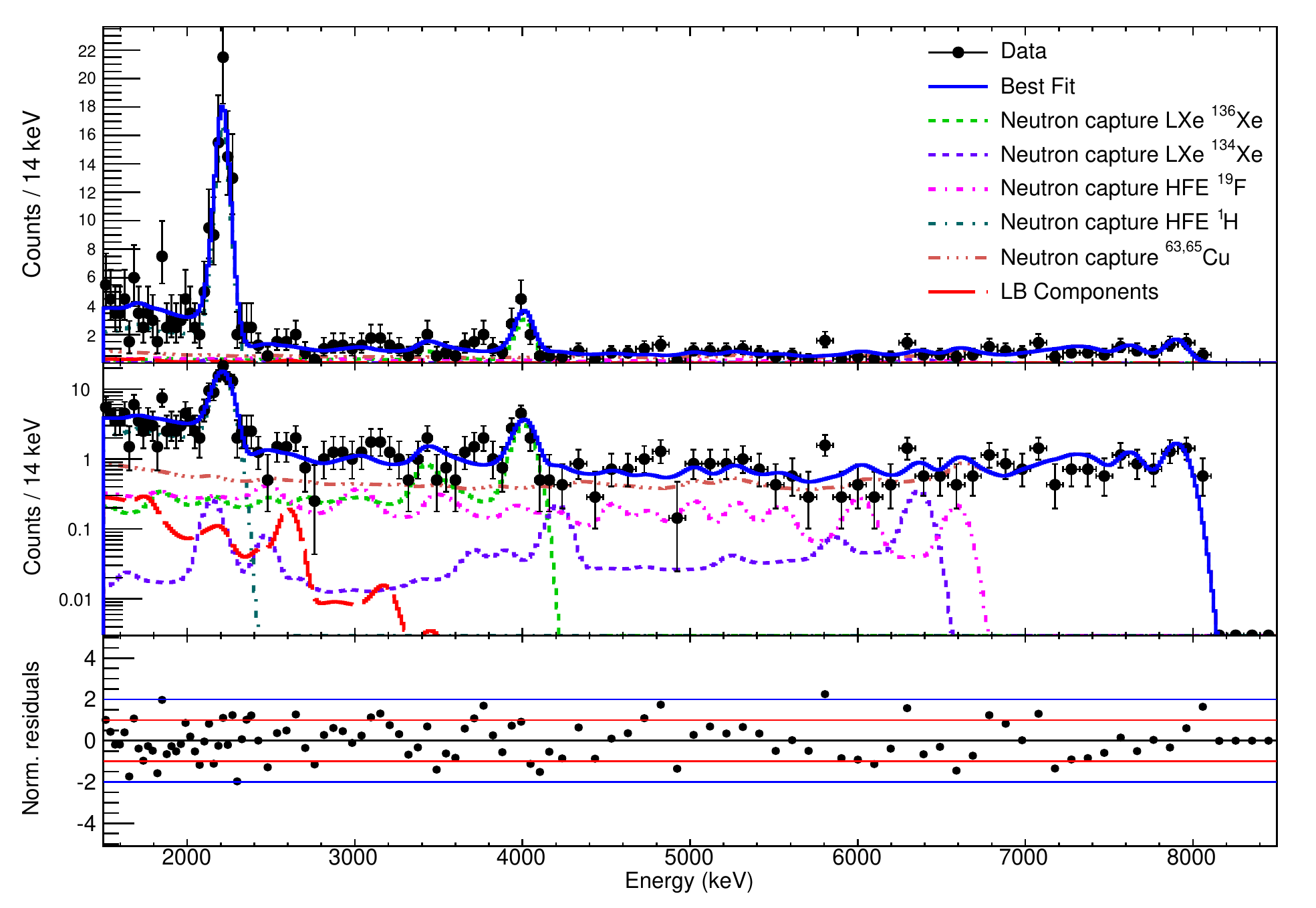}
	\caption[MS Fit] {MS spectrum and the veto-tagged data fit.  The SS
	spectrum and standoff distances were fit simultaneously with this.  The different 
	panels show results on
	a linear scale, logarithmic scale, and residuals, from top to bottom.  The 
	energy-binned fit is
	performed with 14~keV bins, though the binning shown here 
	is modified and non-uniform to preserve spectral
	details while avoiding sparsely populated bins.
	}
	\label{fig:ms_fit}
	\end{center}
\end{figure}

\begin{figure}[htpb]
\begin{minipage}{0.495\linewidth}
	\begin{center}
	\includegraphics[keepaspectratio=true,width=3.26in, trim = 0mm 0mm 0mm 0mm]
	{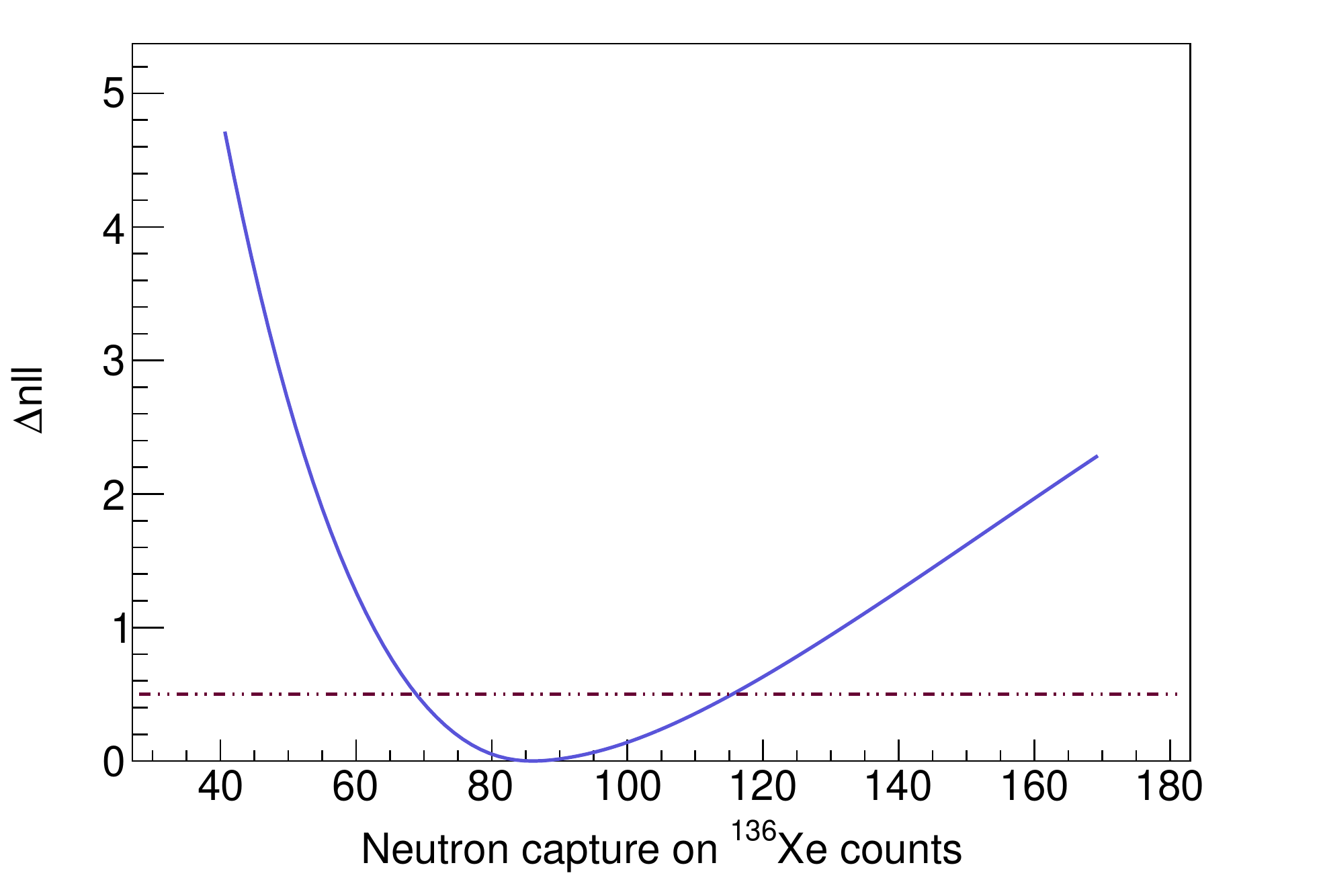}
	\end{center}
\end{minipage}
\begin{minipage}{0.495\linewidth}
	\begin{center}
	\includegraphics[keepaspectratio=true,width=3.26in, trim = 0mm 0mm 0mm 0mm]
	{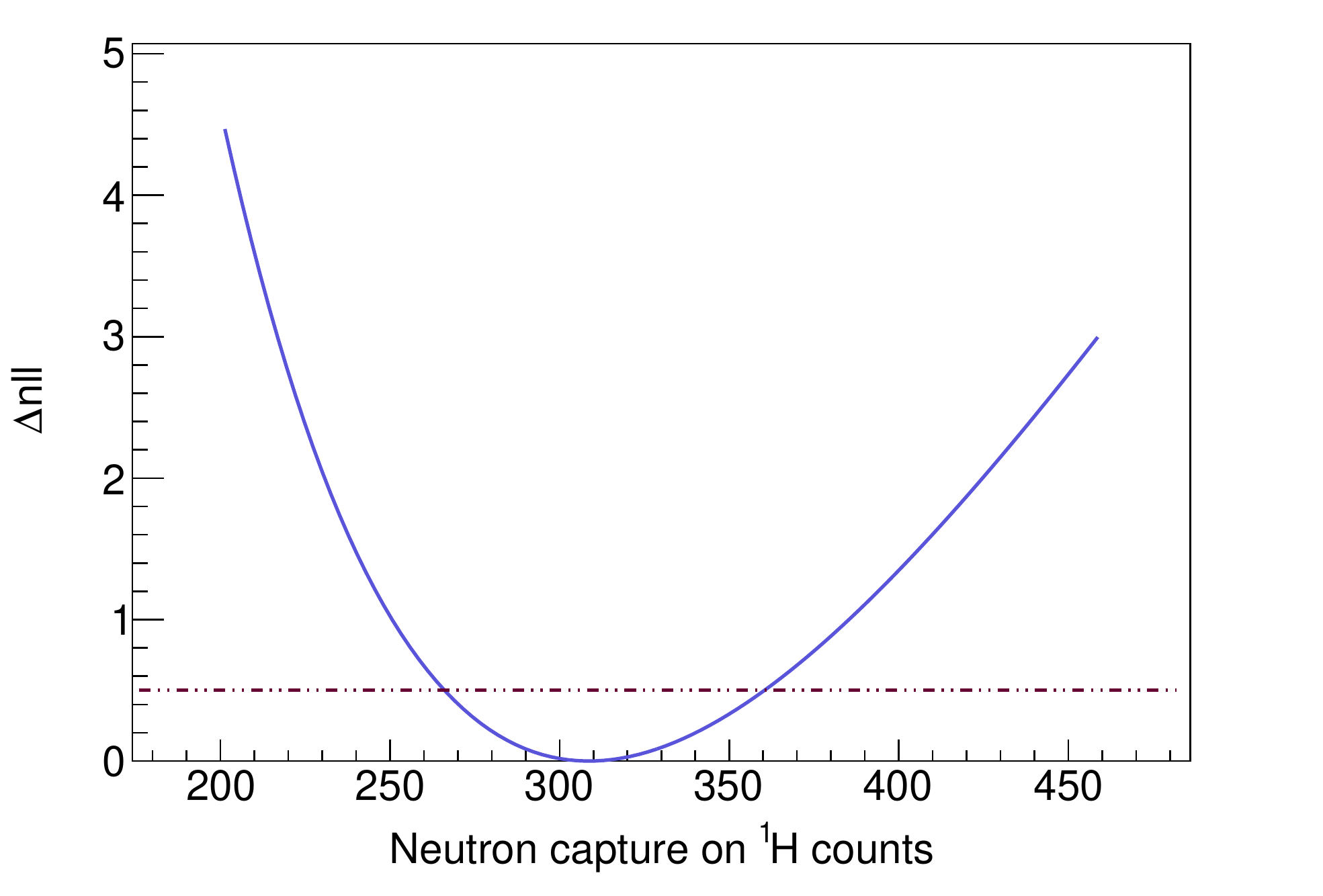}
	\end{center}
\end{minipage}
\caption[Profiles over\otsx and $^{1}\mathrm{H}$]
{Profile likelihood for various values of fitted capture counts on 
\otsx (left) and $^{1}\mathrm{H}$ (right).  The vertical axis is the difference in the fit
negative log likelihood (nll) between a particular number of counts and the best-fit value.  
A dashed horizontal line represents the 
$1~\sigma$ confidence level, assuming the validity of Wilks' 
theorem~\cite{Wilks:1938dza, Cowan:1998ji}.
The resulting fits yielded $86.0^{+29.4}_{-17.2}$ for \otsx and $309^{+52}_{-43}$
for $^{1}\mathrm{H}$ at $1~\sigma$.}
\label{fig:profiles}
\end{figure}

The MS energy spectrum and PDFs fitted to the data are shown in figure~\ref{fig:ms_fit}.
To obtain a final number of counts with uncertainties, 
profile-likelihood scans were performed over the number of counts
in each of these PDFs.
Examples are shown
in figure~\ref{fig:profiles}.

The validity of our fit model was demonstrated by the small deviations of the SS event fractions
and degenerate PDF fractions from the values determined from simulation.  
The SS fractions for the PDFs varied by
between $-0.15\%$ and $-0.49\%$ from their central values, small changes relative to the 5.3\%
constraint.  The degenerate PDF ratios (20\% constraint) typically varied by $\sim$1\%,
with the maximum deviations $-4.64\%$ for the $^{63}\mathrm{Cu}$~TPC vessel and $+4.24\%$
for the innermost shell of HFE-7000~$^{1}\mathrm{H}$.  If there were 
problems with the capture model, such as significant unaccounted-for PDFs, 
larger deviations would be expected.

The various efficiencies and uncertainties for capture on \otsx and $^{1}\mathrm{H}$ are
listed in table~\ref{tbl:veto_tagged_data_sys}.  Several of these efficiencies
are highly affected by pile-up and detector inefficiencies related to muon-induced
events.  As these efficiencies are unique to this veto-tagged data analysis,
details for their evaluation are presented in appendix~\ref{sec:systematics}.

\begin{table}[htbp]
\begin{center}
\begin{tabular}{|l|c|c|c|c|}
\hline
\multicolumn{1}{|C{4cm}|}{Systematic} & \multicolumn{1}{C{2cm}|}{$^{1}\mathrm{H}$ efficiency} & 
\multicolumn{1}{C{2cm}|}{$^{1}\mathrm{H}$ uncertainty} & \multicolumn{1}{C{2cm}|}{\otsx efficiency}
 & \multicolumn{1}{C{2cm}|}{\otsx uncertainty} \\
\hline
Normalization constraint & -- & 10.5\% & -- & 10.5\%\\
~~~---~Noise tagging & -- & 0.4\% & -- & 0.4\%\\
~~~---~3D reconstruction & -- & 4.3\% & -- & 4.3\% \\
~~~---~Overall efficiency & -- & 9.6\% & -- & 9.6\% \\
\hline
Capture cascade & -- & -- & -- & 0.9\% \\
Multi-$\gamma$ efficiency & -- & -- & -- & 8.4\% \\
Capture position distribution & -- & 7.7\% & -- & 1.8\% \\
Veto panel trigger & $95.7\%$ & 1.7\% & $96.0\%$ & 1.7\% \\
Veto time window & $98.7\%$ & 2.5\% & $98.2\%$ & 4.8\% \\
Muon-capture pile-up & $94.7\%$ & 4.8\% & $88.8\%$ & 11.2\% \\
Drift time cut & $85.2\%$ & 7.1\% & $76.7\%$ & 11.7\% \\
Frame-edge effects & $97.1\%$ & 1.6\% & $97.1\%$ & 1.6\% \\
\hline
Total systematic efficiency & $74.2\%$ & 12.3\% & $62.4\%$ & 19.1\% \\
\hline
Single capture efficiency & 0.126\% & -- & 29.6\% & -- \\

\hline
\end{tabular}
\caption{\label{tbl:veto_tagged_data_sys} Summary of veto-tagged data efficiencies and
systematic uncertainties.  All uncertainties are added in quadrature
to arrive at the total uncertainty.
The combination of efficiencies for different shells of HFE-7000 is 
complicated by the varying number of captures on each shell, so the total efficiency
and uncertainty for $^{1}\mathrm{H}$
are not simply related to the individually evaluated efficiencies and uncertainties.
Noise tagging, 3D reconstruction and overall efficiency are all included in the normalization
constraint in the fit; all other terms are applied after fitting.
The systematic efficiency is due entirely
to effects associated with the veto-tagged nature of the data and proximity to other signals.
Single capture efficiency is the efficiency for an isolated veto-tagged capture 
to produce a signal passing the analysis cuts, in the absence of these systematic effects.}
\end{center}
\end{table}

\subsection{Capture rates in data and simulation}

The best-fit results for the number of counts above 1500~keV in each of the PDFs are shown in 
table~\ref{tbl:fit_results}, along with the extrapolation to total number of captures.
The expected numbers of captures from the FLUKA and Geant4 simulations are also included.

This comparison between captures determined with our models alone (without data input)
and measured captures allows
us to evaluate the simulations.  
Based on the rate of capture on $^{1}\mathrm{H}$ (with the smallest measurement uncertainty), 
FLUKA and Geant4 are over-estimating neutron capture by $45^{+33}_{-25}\%$ and $3^{+23}_{-18}\%$, respectively.
Using capture on \otsx to evaluate this, the over-estimates go to $19^{+45}_{-34}\%$ and 
$30^{+49}_{-37}$\%, respectively.  It is not clear whether this is due to differences
in simulated neutron production rates or capture cross-sections.
This level of discrepancy ($\sim$40\%) is common for neutron production/transport 
simulations~\cite{Abe:2009aa, Reichhart:2013xkd, Marino:2007ti, Bellini:2013pxa}.

As the EXO-200 detector is a complex system with large masses of copper,
enriched xenon, lead, and HFE-7000, it is difficult to identify any particular process which is
not being simulated exactly.  Despite the discrepancies and significant measurement uncertainties,
this is a useful validation for both simulation packages, demonstrating reasonable
performance with relevant neutron production
and capture targets, such as enriched xenon.

\begin{table}[htbp]
\renewcommand*\arraystretch{1.2}
\begin{center}
\begin{tabular}{|c|c|c|c|c|}
\hline
\multicolumn{1}{|C{3.0cm}|}{PDF} &
\multicolumn{1}{C{1.9cm}|}{Best-fit \newline counts} &
\multicolumn{1}{C{2.8cm}|}{Measured \newline captures/year} &
\multicolumn{1}{C{2.4cm}|}{Geant4 \newline captures/year} &
\multicolumn{1}{C{2.4cm}|}{FLUKA \newline captures/year} \\
\hline
Low-background & $35.9\pm11.7$ & -- & -- & -- \\
Copper capture & $306^{+70}_{-54}$ &  $4.23^{+1.30}_{-1.15}\times10^{5}$ & $4.69\times10^{5}$ & $5.78\times10^{5}$\\ 
$^{19}\mathrm{F}$ capture & $87.1^{+38.0}_{-35.1}$ & $2.85^{+1.29}_{-1.20}\times10^{4}$ & $2.75\times10^{4}$ & $2.45\times10^{4}$\\ 
$^{1}\mathrm{H}$ capture & $309^{+52}_{-43}$ &  $2.41^{+0.50}_{-0.44}\times10^{5}$ & $2.48\times10^{5}$ & $3.50\times10^{5}$\\ 
$^{134}\mathrm{Xe}$ capture & $20.1^{+18.2}_{-13.9}$ & $72.9^{+67.1}_{-52.0}$ & 116 & 120\\ 
$^{136}\mathrm{Xe}$ capture & $86.0^{+29.4}_{-17.2}$ & $338^{+132}_{-93}$ & 439 & 403\\ 
\hline
\end{tabular}
\caption{\label{tbl:fit_results} Summary of the results of the veto-tagged fit and
simulations.  The best-fit counts
are the total (SS and MS) number of counts above 1500~keV.
This is based on a
 simultaneous fit in energy and standoff distance for SS and MS, while profiling over
 the number of capture counts.
Final capture rates are for all captures.  Component masses can be found 
in table~\ref{tab:componentmasses}. 
Only captures on $^{1}\mathrm{H}$ and \otsx are considered to have fully validated 
systematics, as other capture PDFs have high uncertainties for capture cascade and/or
capture timing efficiency.  Measured capture rates are provided for these more uncertain
PDFs for completeness.  Uncertainties on simulation predictions are not listed, but
are dominated by the 3.5\% uncertainty on measured muon flux.
}
\end{center}
\end{table}

The \znbb low-background fit (based on the \znbb analysis~\cite{Albert:2014awa}) measures a 
$^{137}\mathrm{Xe}$ decay rate of $374^{+135}_{-93}$ decays per year.  As seen in 
table~\ref{tbl:fit_results}, this is consistent,
within uncertainties, with the measured production rate from capture $\gamma$s ($338^{+132}_{-93}$), 
and with the production rates computed
from the FLUKA and Geant4 simulations (403 and 439, respectively).
The capture detection 
efficiencies for nuclides other than \otsx and $^{1}\mathrm{H}$ have larger, 
difficult to evaluate uncertainties due to much
higher resonant neutron capture rates, capture cascade uncertainties, and/or lower statistics, 
so careful comparison to simulation is not attempted.

A search of low-background data showed no evidence for significant \otsx neutron capture
out of coincidence with the muon veto.  At the $1~\sigma$ C.L., no more than $10\%$ of the 
$^{137}\mathrm{Xe}$ production is due
to non-cosmogenic sources, and the best-fit rate is nearly zero.  
This is consistent with simulations (using SOURCES4A~\cite{SOURCES}, FLUKA, and Geant4) 
of expected $(\alpha,n)$ and spontaneous fission
neutrons coming from the salt surrounding the detector cavern.  It is
notable that the copper cryostat and TPC, along with the HFE-7000 bath, manage to very
effectively shield the xenon from radioactive sources of thermal neutrons, leaving only
cosmogenic neutrons to produce significant backgrounds.

Identification of neutron
captures can be used to remove $^{137}\mathrm{Xe}$ decays from the \znbb analysis data set.  
Such a veto would last several half-lives, and could be targeted to
only reject signals in parts of the TPC where the capture was observed.  This veto will
be optimized and implemented in future \znbb analyses using \otsx.

\section{\label{sec:conclusions}Conclusions}
We have used the EXO-200 TPC to perform a measurement of the muon flux  at WIPP,
finding a vertical intensity of 
$I_{\mathrm{v}} = 2.97^{+0.14}_{-0.13}~\mathrm{(sys)}\pm0.02~\mathrm{(stat)}\times 10^{-7} \mathrm{cm^{-2}~s^{-1}~sr^{-1}}$, and
a flux of $\Phi = 4.07\pm0.14~\mathrm{(sys)}\pm0.03~\mathrm{(stat)}\times 10^{-7} \mathrm{cm^{-2}~s^{-1}}$.
We used this
measurement to normalize MC model calculations for cosmogenic neutron production, transport and
capture, and cosmogenic radionuclide production.  
The only cosmogenic radionuclide found to have a significant contribution to \znbb backgrounds
is $^{137}\mathrm{Xe}$.  Other possible cosmogenic backgrounds include several radioisotopes
of iodine, but these were found to contribute much less than 0.1 counts per year to the EXO-200
ROI for \znbb, based on MC simulations.

We also directly measured cosmogenic neutron capture by looking at the capture
cascades in data coincident with veto triggers.  This technique yielded an estimate of 
$338^{+132}_{-93}$ captures on \otsx per year.  This is in 
agreement with the observed $^{137}\mathrm{Xe}$ decay rate and the estimates from FLUKA
and Geant4.  
In general, both simulations appear to be reasonably accurate. The FLUKA estimate for 
neutron capture on $^{1}\mathrm{H}$ is $45^{+33}_{-25}\%$ higher than the measured rate, in
tension by more than $2~\sigma$, while the Geant4 estimates agree within uncertainties.
Discrepancies of the observed magnitude are common for neutron production and transport simulations.
The technique of identifying neutron captures using veto-tagged data, demonstrated here, 
will be used in future analyses to reject 
$^{137}\mathrm{Xe}$ decays from the \znbb low-background dataset.

\acknowledgments

EXO-200 is supported by DOE and NSF in the United
States, NSERC in Canada, SNF in Switzerland, IBS in
Korea, RFBR (14-02-00675) in Russia, DFG Cluster 
of Excellence ``Universe'' in Germany, and CAS-IHEP Fund and
ISTCP (2015DFG02000) in China. EXO-200 data
analysis and simulation uses resources of the National
Energy Research Scientific Computing Center (NERSC),
which is supported by the Office of Science of the U.S.
Department of Energy under Contract No. DE-AC02-05CH11231.
We thank Milan~Krti\u{c}ka for his assistance with DICEBOX.
The collaboration gratefully acknowledges the KARMEN Collaboration for supplying the
veto detectors, and the WIPP for their hospitality.

\appendix

\section{\label{sec:expanded_table}Expanded table of cosmogenic nuclides of interest}
  Table~\ref{tbl:expanded_isotopes} includes the nuclides expected to have the largest contribution
to the \znbb energy ROI. Several other nuclides are also included as they may be
of interest.

\begin{table}[htbp]
\centering
\begin{tabular}{|c c c c c c|}
\hline
  \multicolumn{1}{|C{2.5cm}}{Region/ \newline Nuclide} & \multicolumn{1}{C{1.1cm}}{Half-\newline life}
   & \multicolumn{1}{C{1.75cm}}{Total energy (keV)} & \multicolumn{1}{C{1.1cm}}{Decay mode}
   & \multicolumn{1}{C{2.2cm}}{Geant4 rate (yr$^{-1}$)} & \multicolumn{1}{C{2.2cm}|}{FLUKA rate (yr$^{-1}$)} \\
\hline
Cryostat~$^{62}\mathrm{Co}$ & 1.5~m & 5315 & $\beta+\gamma$ & $97.1\pm3.7$ & $92.9\pm4.7$ \\
Cryostat~$^{60}\mathrm{Cu}$ & 24~m & 6127 & $\mathrm{EC}/\beta+\gamma$ & $82.2\pm3.2$ & $250\pm10$ \\\hline
HFE~$^{8}\mathrm{Li}$ & 0.84~s & 16004 & $\beta$ & $64.1\pm2.5$ & $354\pm14$ \\
HFE~$^{9}\mathrm{Li}$ & 0.18~ms & 13606 & $\beta$ & $8.7\pm0.5$ & $63.0\pm3.6$ \\
HFE~$^{11}\mathrm{Be}$ & 13.8~s & 11506 & $\beta+\gamma$ & $2.3\pm0.2$ & $23.7\pm1.9$ \\
HFE~$^{8}\mathrm{B}$ & 0.77~s & 17979 & $\mathrm{EC}/\beta$ & $5.1\pm0.4$ & $50.3\pm3.1$ \\
HFE~$^{12}\mathrm{B}$ & 20~ms & 13369 & $\beta+\gamma$ & $107\pm4$ & $368\pm15$ \\
HFE~$^{13}\mathrm{B}$ & 17~ms & 13437 & $\beta+\gamma$ & $17.6\pm0.8$ & $79.6\pm4.2$ \\
HFE~$^{15}\mathrm{C}$ & 2.4~s & 9772 & $\beta+\gamma$ & $146\pm5$ & $36.3\pm2.5$ \\
HFE~$^{17}\mathrm{C}$ & 0.19~s & 13166 & $\beta+\gamma$ & $69.6\pm2.7$ & 0.00 \\
HFE~$^{12}\mathrm{N}$ & 11~ms & 17338 & $\mathrm{EC}/\beta+\gamma$ & $6.1\pm0.4$ & $5.5\pm0.8$ \\
HFE~$^{16}\mathrm{N}$ & 7.1~s & 10420 & $\beta+\gamma$ & $2380\pm80$ & $2910\pm110$ \\
HFE~$^{18}\mathrm{N}$ & 0.62~s & 13899 & $\beta+\gamma$ & $9.9\pm0.5$ & $0.8\pm0.3$ \\
HFE~$^{14}\mathrm{O}$ & 71~s & 5143 & $\mathrm{EC}/\beta+\gamma$ & $38.6\pm1.6$ & $19.3\pm1.7$ \\
HFE~$^{19}\mathrm{O}$ & 27~s & 4821 & $\beta+\gamma$ & $615\pm22$ & $691\pm26$ \\
HFE~$^{20}\mathrm{F}$ & 11~s & 7025 & $\beta+\gamma$ & $27490\pm980$ & $23030\pm820$ \\\hline
TPC vessel~$^{56}\mathrm{Co}$ & 77~d & 4566 & $\mathrm{EC}/\beta+\gamma$ & $0.5\pm0.1$ & $1.3\pm0.4$ \\
TPC vessel~$^{60}\mathrm{Co}$ & 5.3~y & 2824 & $\beta+\gamma$ & $2.6\pm0.2$ & $2.9\pm0.6$ \\
TPC vessel~$^{62}\mathrm{Cu}$ & 9.7~m & 3948 & $\mathrm{EC}/\beta+\gamma$ & $71.7\pm2.8$ & $76.9\pm4.1$ \\\hline
TPC LXe~$^{8}\mathrm{Li}$ & 0.84~s & 16004 & $\beta$ & $0.02\pm0.02$ & $1.1\pm0.4$ \\
TPC LXe~$^{108}\mathrm{In}$ & 58~m & 5157 & $\mathrm{EC}/\beta+\gamma$ & $0.14\pm0.05$ & $0.25\pm0.18$ \\
TPC LXe~$^{120}\mathrm{I}$ & 81~m & 5615 & $\mathrm{EC}/\beta+\gamma$ & $0.4\pm0.1$ & $2.8\pm0.6$ \\
TPC LXe~$^{124}\mathrm{I}$ & 4.2~d & 3160 & $\mathrm{EC}/\beta+\gamma$ & $2.5\pm0.2$ & $8.4\pm1.1$ \\
TPC LXe~$^{130}\mathrm{I}$ & 12~h & 2949 & $\beta+\gamma$ & $7.3\pm0.4$ & $21.6\pm1.8$ \\
TPC LXe~$^{132}\mathrm{I}$ & 2.3~h & 3577 & $\beta+\gamma$ & $7.7\pm0.5$ & $22.2\pm1.8$ \\
TPC LXe~$^{134}\mathrm{I}$ & 53~m & 4175 & $\beta+\gamma$ & $7.3\pm0.4$ & $20.4\pm1.7$ \\
TPC LXe~$^{135}\mathrm{I}$ & 6.6~h & 2648 & $\beta+\gamma$ & $8.6\pm0.5$ & $21.6\pm1.8$ \\
TPC LXe~$^{136}\mathrm{I}$ & 83~s & 6926 & $\beta+\gamma$ & $2.1\pm0.2$ & $1.7\pm0.5$ \\
TPC LXe~$^{121}\mathrm{Xe}$ & 40~m & 3745 & $\mathrm{EC}/\beta+\gamma$ & $0.6\pm0.1$ & $1.0\pm0.4$ \\
TPC LXe~$^{135}\mathrm{Xe}$ & 9.1~h & 1151 & $\beta+\gamma$ & $1110\pm40$ & $1060\pm40$ \\
TPC LXe~$^{137}\mathrm{Xe}$ & 3.8~m & 4173 & $\beta+\gamma$ & $439\pm17$ &  $403\pm16$\\
\hline
\end{tabular}
\caption{\label{tbl:expanded_isotopes} Expanded table of cosmogenic radionuclides of interest.  
The half-lives, decay mode, and total decay energy are 
listed~\cite{Firestone:391553}.  
The Geant4 and FLUKA columns
give the estimated production rate for the listed nuclide in atoms per year.
Uncertainties are dominated by muon flux
and MC statistics.
}
\end{table}

\section{\label{sec:systematics}Veto-tagged data systematic effects}

The normalization constraint (see published \tnbb analysis~\cite{Albert:2013gpz} for 
implementation details and theory) 
was applied to all PDF components and serves to incorporate an overall efficiency uncertainty
into the profile-likelihood fit.
This 10.5\% uncertainty is composed of systematic errors
 added in quadrature for noise tagging, 3D reconstruction efficiency, 
and overall efficiency.  The efficiency uncertainties came from averaged rate differences 
between $^{60}\mathrm{Co}$, 
$^{226}\mathrm{Ra}$, and $^{228}\mathrm{Th}$ source data and MC with sources deployed at
several positions
near the TPC.  The noise tag uncertainty was evaluated by reviewing noise-tagged
data events by visual inspection, looking for falsely noise-tagged valid events.

The 5.3\% SS fraction constraint was based on the maximum deviation of the SS fraction 
determined with
source data and MC among all tested sources and deployment positions.  This maximum
deviation was observed in $^{228}\mathrm{Th}$ data with source near the cathode.

All remaining systematic uncertainties were applied to the measurements
after fitting, with individual uncertainties added in quadrature.
These were evaluated for each capture PDF.

An additional 8\% uncertainty was added separately to the \otsx neutron capture detection 
efficiency
to bring the total efficiency uncertainty to 12.5\%, the maximum deviation observed in
$^{60}\mathrm{Co}$ sum-peak data.  This was
motivated by the fact that $^{136}\mathrm{Xe}(n,\gamma)$ is a multi-$\gamma$ process, and may have
systematic effects not accounted for in single-$\gamma$ source studies. A 
2.5\% uncertainty is added in quadrature to this to account for the effects of an observed
slight shift in the energy scale for the $^{136}\mathrm{Xe}(n,\gamma)$ PDF, likely due to
its multi-$\gamma$ nature.
An additional cascade model uncertainty was applied for the \otsx capture as well.  The
cascade model uncertainty evaluation is described in section~\ref{sec:ncap_gammas}.

The efficiency to detect captures depends strongly on their position distribution.
The relative number of captures calculated by FLUKA on each degenerate component (such as the 
concentric shells of HFE-7000) was assumed to be our most accurate estimate.
To account for uncertainties in the modeling of capture positions among degenerate components, 
differences in average capture $\gamma$ detection efficiency resulting from 
the capture position distributions obtained from 
FLUKA and Geant4 were treated as a systematic uncertainty.  Deviations
in the fitted degenerate PDF ratios from their simulation values were evaluated to add to
this position uncertainty.
To account for non-uniform captures within each component (within a single shell of HFE,
for example), the Geant4 capture positions were used to
determine a small correction, which was applied for all efficiencies.

The probability for each type of capture to be coincident with a veto panel trigger was estimated
from FLUKA simulation.  These raw simulated efficiencies were scaled by a factor of 0.983, 
necessary to bring the 97.99\% FLUKA efficiency for TPC muons to be tagged by veto panels 
in line with the $96.35\pm0.11\%$ value determined from the full
data set.  This need for a correction was attributed to the lack of response modeling of the
veto detectors in simulation.  Although corrected for, we treated this efficiency
scaling as an additional uncertainty.

\begin{figure}[htpb]
\begin{minipage}{0.495\linewidth}
	\begin{center}
	\includegraphics[keepaspectratio=true,width=\linewidth, trim = 0mm 0mm 13mm 0mm]
	{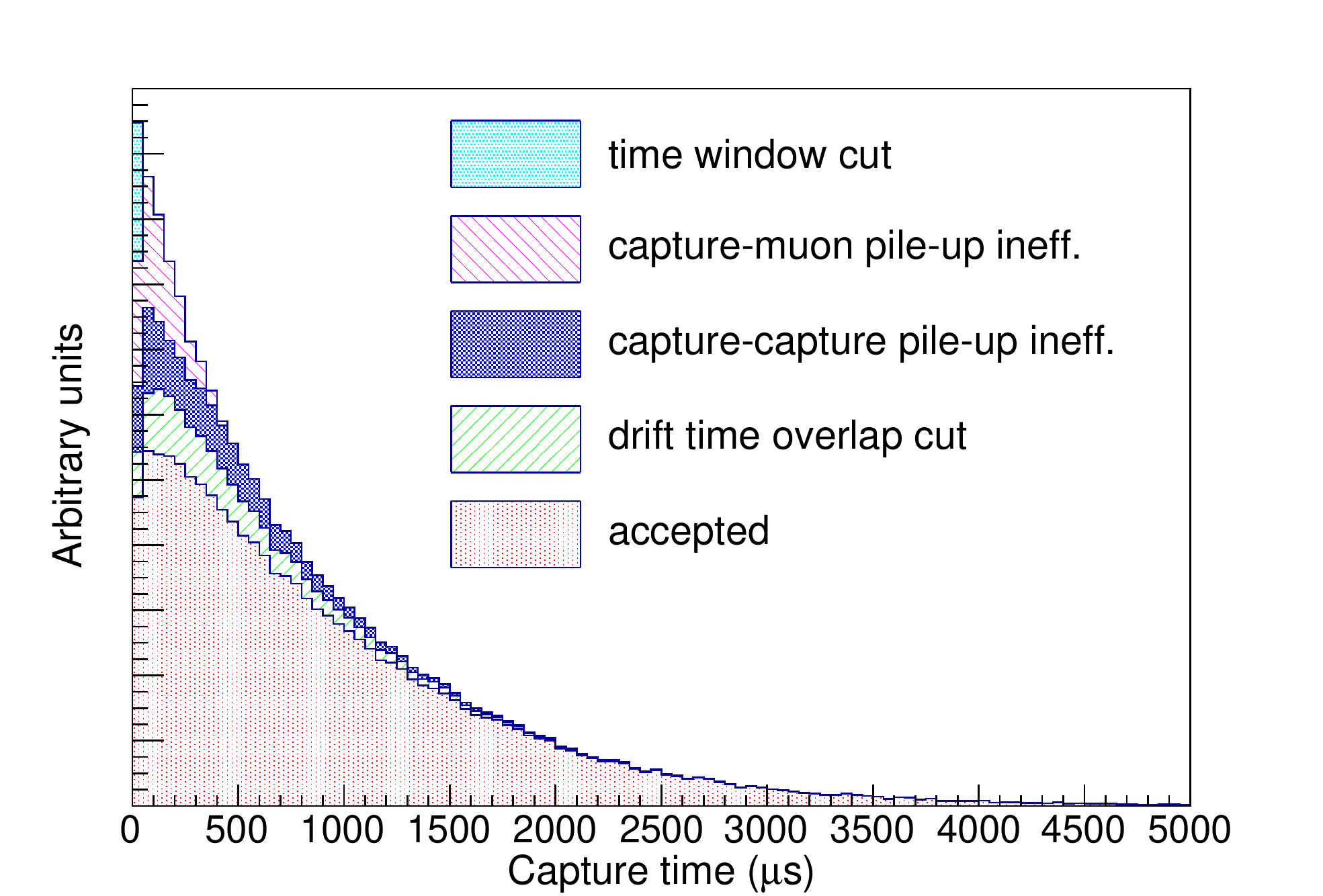}
	\end{center}
\end{minipage}
\begin{minipage}{0.495\linewidth}
	\begin{center}
	\includegraphics[keepaspectratio=true,width=\linewidth, trim = 0mm 0mm 13mm 0mm]
	{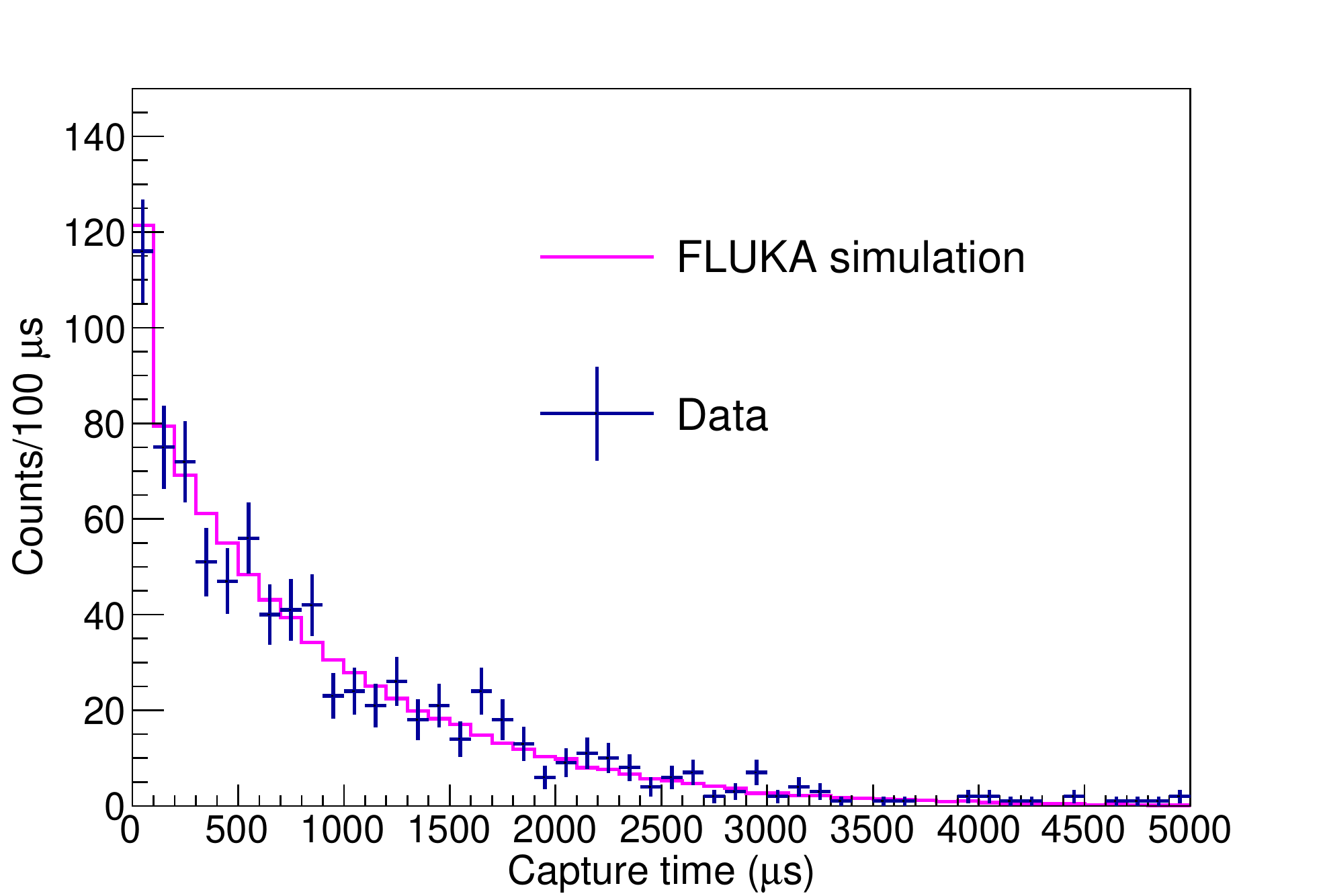}	
	\end{center}
\end{minipage}
\caption[FLUKA simulation of capture times]
{ (left) Distribution of the observed capture times (time since muon) for 
neutrons on $^{1}\mathrm{H}$ in HFE-7000, as estimated from FLUKA
simulation.  Each of the timing cuts and efficiencies (except the frame edge efficiency)
are applied, reducing the total efficiency for observing these captures.  (right) FLUKA
distribution for observed capture times for all veto-tagged events, after 
time-dependent efficiencies are applied.  Data from veto-tagged events are overlaid
in blue, and the number of MC counts is scaled to match the data.  
In both panels, the FLUKA capture times have been scaled by 1.05.
}
\label{fig:fluka_timing}
\end{figure}

Several contributions to the total efficiency depend on the time 
distribution of neutron captures.  The following effects
were considered: capture time window (10~$\mu$s - 5~ms), reduced reconstruction efficiency
following a TPC muon, reduced reconstruction efficiency following another capture event,
drift time cut (119~$\mu$s between scintillation events), and reduced efficiency
at the edges of the data frame.  Models for these efficiencies were developed based on 
radioactive source data and other data.  The capture efficiencies were estimated by applying these models
(with the exception of the frame edge efficiency, applied separately) 
to the FLUKA simulated capture events.  We chose the FLUKA
simulation for this because it replicated the observed capture time distribution better
than Geant4.  Agreement between the FLUKA expected capture times and data was tested by
fitting capture times with an exponential and comparing data and MC.  Optimal agreement was
found by scaling the FLUKA capture times by a factor of 1.05, so this is applied for all
FLUKA capture time studies.
Plots of the FLUKA capture time distribution and the aforementioned efficiencies
are shown in figure~\ref{fig:fluka_timing}.

The capture time window cut and drift time cuts were applied to the FLUKA simulation in a
straightforward manner.
The muon-capture pile-up inefficiency
was evaluated and parameterized based on veto-tagged data.  Our reconstruction
algorithms are optimized for isolated scintillation clusters, and are less sensitive to signals
occurring within several hundred $\mu$s after the large muon signal.  These capture signals can still
be identified with simple thresholding technique, at the expense of energy and timing
precision.  We search for missed scintillation clusters using this technique, thus finding the
fraction of captures missed as a function of time since TPC muon.  

\begin{figure}[htpb]
\begin{minipage}{0.495\linewidth}
	\begin{center}
	\includegraphics[keepaspectratio=true,width=\linewidth, trim = 6mm 0mm 7mm 0mm]
	{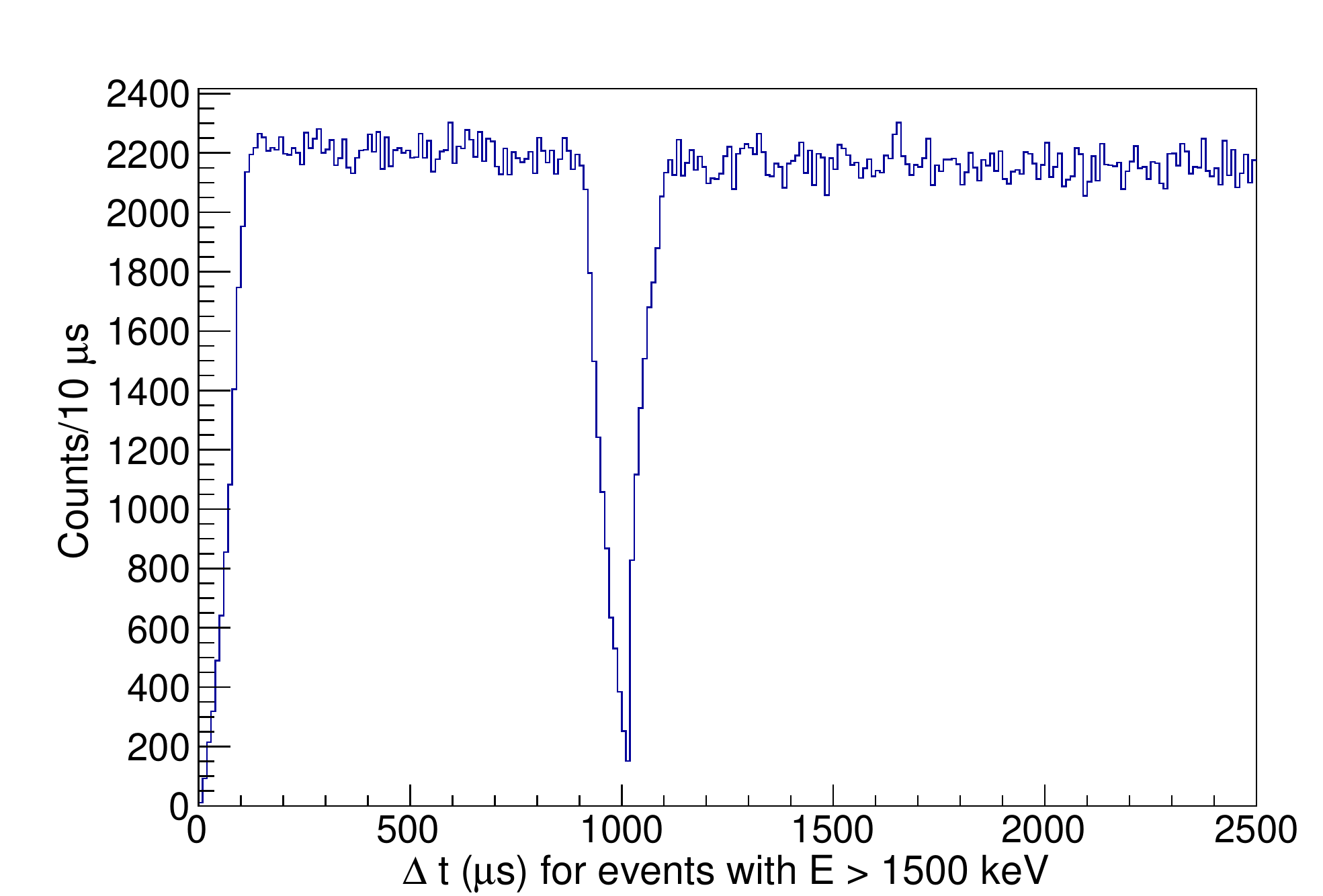}
	\end{center}
\end{minipage}
\begin{minipage}{0.495\linewidth}
	\begin{center}
	\includegraphics[keepaspectratio=true,width=\linewidth, trim = -1mm 0mm 14mm 0mm]
	{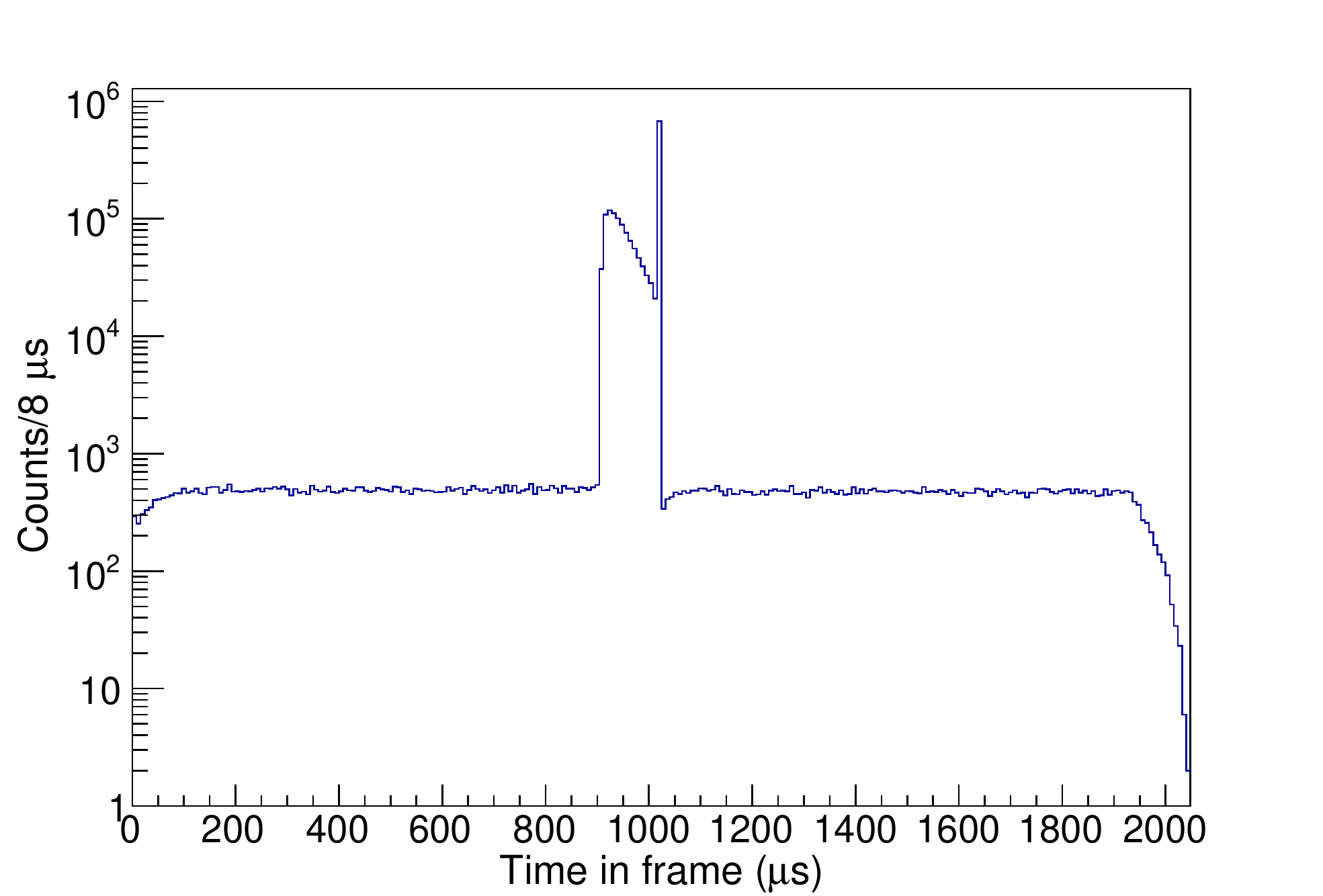}
	\end{center}
\end{minipage}
\caption[Source studies for timing]
{Examples of source studies for timing-based efficiency.  On the left is the distribution of
events as a function of $\Delta t$, time since the previous event, for data from a 
$^{228}\mathrm{Th}$ source deployed at a position near the 
cathode (designated ``S5'').  The deficit at short times was fitted, and the resulting
efficiency as a function of $\Delta t$ was applied to the FLUKA MC to incorporate this effect.  
The scintillation cluster-finding algorithm is optimized for isolated clusters, and does not
perform well when scintillation events occur in quick succession.  
The deficit near 1024~$\mu$s is due to inefficiencies for reconstructing signals at the very
end or very beginning of a data frame, and this effect is explored in more detail
in the right panel.  The right
panel shows the distribution of scintillation cluster times in the 2~ms data frame for
$^{226}\mathrm{Ra}$ data at S5.  The DAQ software records the data such that the trigger
is set to 1024~$\mu$s, so the sharp peak and broad peak in the feature near the frame center are due
to triggers on scintillation light and charge, respectively. 
The flat part of the distribution is due to pile-up.  The deficits at the beginning and
end of the frame are fitted and used to estimate rates of event loss due to frame-edge
effects.
Several different source positions needed to be used as the 
frame-edge efficiency is highly dependent on the time for charge deposits to drift to the
anode.  In both cases, only events with $E > 1500~\mathrm{keV}$ are considered.
}
\label{fig:source_timing}
\end{figure}

The reconstruction efficiency dips for any scintillation clusters too near each
other in time (capture-capture pile-up), and for scintillation clusters at the beginning 
and end of the 2048~$\mu$s data frame (frame-edge effects).  These
were quantified using source data.  The deficit in the time between events ($\Delta t$) 
distribution at small
$\Delta t$ in source data provides a measure of signal loss.  The fraction of 
counts at the beginning and end of a frame to be lost in source data was used to quantify the frame edge
timing loss.   See figure~\ref{fig:source_timing} for examples.
All events with scintillation clusters within 120~$\mu$s of the end of a frame are not used for analysis
(as charge clusters may be lost if their collection signal waveforms are not completed by
the end of the frame), 
but we used the number of observed veto-tagged counts
in this time to extrapolate, using the source data efficiency, to the number of events
lost at the end of data frames.  The number of events lost at the beginning of data frames
was related to the number lost at the end based on the source data studies.
This extrapolation yielded a $97.1\pm1.6\%$ data frame efficiency, which was applied to all PDFs.

After these reconstruction efficiencies were introduced to the FLUKA MC, calculated efficiencies
for each PDF were evaluated.  All timing efficiencies came directly from this modified FLUKA
MC, except
for the drift time cut efficiency and frame-edge efficiency.  The capture-capture pile-up
efficiency has no effect on total efficiency after the drift time cut is applied.
For the drift time cut, FLUKA estimated an overall efficiency of 89.5\% for veto-tagged events, 
while data showed
a 94.6\% efficiency (after correction for the capture-capture pile-up inefficiency).  
To account for this discrepancy, a scale factor for this inefficiency of $0.75\pm0.37$
was applied to the inefficiency for each PDF, yielding an adjusted overall drift time cut efficiency of
$92.1\pm3.9\%$, in agreement within uncertainties with the efficiencies from both data and MC.
The capture-capture pile-up efficiency (average 89.9\% for all data) was then added in,
yielding a final average efficiency of $82.8\pm6.4\%$.

The uncertainty for the time window efficiency was set to the maximum difference between
Geant4 and FLUKA calculations (with and without time scaling factors).  
The uncertainty for TPC muon-induced inefficiency was 
conservatively set to the entirety of the correction, and the frame edge efficiency uncertainty 
was based on differences between different source position results.

\bibliographystyle{./JHEP.bst}
\bibliography{cosmogenic-backgrounds-jcap}

\providecommand{\href}[2]{#2}\begingroup\raggedright\begin{thebibliography}{10}

\bibitem{Gando:2012zm}
{\bf KamLAND-Zen} Collaboration, A.~Gando et~al., {\it {Limit on Neutrinoless
  $\beta\beta$ Decay of $^{136}$Xe from the First Phase of KamLAND-Zen and
  Comparison with the Positive Claim in $^{76}$Ge}},  {\em Phys. Rev. Lett.}
  {\bf 110} (2013) 062502, [\href{http://arxiv.org/abs/1211.3863}{{\tt
  arXiv:1211.3863}}].

\bibitem{Albert:2014awa}
{\bf EXO-200} Collaboration, J.~B. Albert et~al., {\it {Search for Majorana
  neutrinos with the first two years of EXO-200 data}},  {\em Nature} {\bf 510}
  (2014) 229, [\href{http://arxiv.org/abs/1402.6956}{{\tt arXiv:1402.6956}}].

\bibitem{Agostini:2013mzu}
{\bf GERDA} Collaboration, M.~Agostini et~al., {\it {Results on Neutrinoless
  Double-$\beta$ Decay of $^{76}$Ge from Phase I of the GERDA Experiment}},
  {\em Phys. Rev. Lett.} {\bf 111} (2013) 122503,
  [\href{http://arxiv.org/abs/1307.4720}{{\tt arXiv:1307.4720}}].

\bibitem{Auger:2012gs}
{\bf {EXO-200}} Collaboration, M.~Auger et~al., {\it {The EXO-200 detector,
  part I: Detector design and construction}},  {\em JINST} {\bf 7} (2012)
  P05010, [\href{http://arxiv.org/abs/1202.2192}{{\tt arXiv:1202.2192}}].

\bibitem{Redshaw:2007un}
M.~Redshaw, E.~Wingfield, J.~McDaniel, and E.~G. Myers, {\it {Mass and
  double-beta-decay Q value of Xe-136}},  {\em Phys. Rev. Lett.} {\bf 98}
  (2007) 053003.

\bibitem{3m}
{3M Novec 7000 Engineered Fluid (HFE-7000), http://www.3m.com}.

\bibitem{Gemmeke:1990ix}
H.~Gemmeke et~al., {\it {The High resolution neutrino calorimeter KARMEN}},
  {\em Nucl. Instrum. Meth.} {\bf A289} (1990) 490.

\bibitem{PhysRevC.92.045504}
{\bf EXO-200} Collaboration, J.~B. Albert et~al., {\it {Measurements of the ion
  fraction and mobility of $\ensuremath{\alpha}\text{-}$ and
  $\ensuremath{\beta}$-decay products in liquid xenon using the EXO-200
  detector}},  {\em Phys. Rev. C} {\bf 92} (2015) 045504,
  [\href{http://arxiv.org/abs/1506.00317}{{\tt arXiv:1506.00317}}].

\bibitem{Albert:2015nta}
{\bf {EXO-200}} Collaboration, J.~B. Albert et~al., {\it {Investigation of
  radioactivity-induced backgrounds in EXO-200}},  {\em Phys. Rev.} {\bf C92}
  (2015) 015503, [\href{http://arxiv.org/abs/1503.06241}{{\tt
  arXiv:1503.06241}}].

\bibitem{Albert:2013gpz}
{\bf EXO-200} Collaboration, J.~B. Albert et~al., {\it {Improved measurement of
  the $2\nu\beta\beta$ half-life of $^{136}$Xe with the EXO-200 detector}},
  {\em Phys. Rev.} {\bf C89} (2014) 015502,
  [\href{http://arxiv.org/abs/1306.6106}{{\tt arXiv:1306.6106}}].

\bibitem{Carlson1969267}
G.~C. Carlson, W.~C. {Schick Jr.}, W.~L. {Talbert Jr.}, and F.~K. Wohn, {\it
  Half-lives of some short-lived mass-separated gaseous fission products and
  their daughters},  {\em Nuclear Physics A} {\bf 125} (1969) 267.

\bibitem{PhysRev.136.B365}
R.~J. Onega and W.~W. Pratt, {\it {Decay of Xe$^{137}$}},  {\em Phys. Rev.}
  {\bf 136} (1964) B365.

\bibitem{Browne20072173}
E.~Browne and J.~K. Tuli, {\it {Nuclear Data Sheets for A = 137}},  {\em
  Nuclear Data Sheets} {\bf 108} (2007) 2173. {Data extracted from the ENSDF
  database, version February 12, 2015, \url{http://www.nndc.bnl.gov/}}.

\bibitem{Allison:2006ve}
J.~Allison et~al., {\it {Geant4 developments and applications}},  {\em IEEE
  Trans. Nucl. Sci.} {\bf 53} (2006) 270.

\bibitem{Agostinelli:2002hh}
{\bf GEANT4} Collaboration, S.~Agostinelli et~al., {\it {GEANT4: A Simulation
  toolkit}},  {\em Nucl. Instrum. Meth.} {\bf A506} (2003) 250.

\bibitem{Agashe:2014kda}
{\bf Particle Data Group} Collaboration, K.~A. Olive et~al., {\it {Review of
  Particle Physics}},  {\em Chin. Phys.} {\bf C38} (2014) 090001.

\bibitem{Miyake:1973qk}
S.~Miyake, {\it {Rapporteur Paper on Muons and Neutrinos}},  in {\em
  {Proceedings of the 13th International Cosmic Ray Conference}}, vol.~5,
  p.~3638, 1973.

\bibitem{Mei:2005gm}
D.~Mei and A.~Hime, {\it {Muon-induced background study for underground
  laboratories}},  {\em Phys. Rev.} {\bf D73} (2006) 053004,
  [\href{http://arxiv.org/abs/astro-ph/0512125}{{\tt astro-ph/0512125}}].

\bibitem{Esch:2004zj}
E.-I. Esch et~al., {\it {The Cosmic ray muon flux at WIPP}},  {\em Nucl.
  Instrum. Meth.} {\bf A538} (2005) 516,
  [\href{http://arxiv.org/abs/astro-ph/0408486}{{\tt astro-ph/0408486}}].

\bibitem{Duda:1972:UHT:361237.361242}
R.~O. Duda and P.~E. Hart, {\it Use of the {H}ough transformation to detect
  lines and curves in pictures},  {\em Commun. ACM} {\bf 15} (1972) 11.

\bibitem{Herrin:2013iq:appxA}
S.~Herrin, {\em {Double Beta Decay in Xenon-136: Measuring the
  Neutrino-Emitting Mode and Searching for Majoron-Emitting Modes}}.
\newblock PhD thesis, Stanford U., Phys. Dept., 2013.
\newblock Appendix A.

\bibitem{Gaisser:1985yw}
T.~Gaisser and T.~Stanev, {\it {Muon Bundles in Underground Detectors}},  {\em
  Nucl. Instrum. Meth.} {\bf A235} (1985) 183.

\bibitem{Cannon:1971jh}
T.~M. Cannon and R.~O. Stenerson, {\it {Multiple muons deep underground}},
  {\em J. Phys.} {\bf A4} (1971) 266.

\bibitem{Cassiday:1973tx}
G.~L. Cassiday, J.~W. Keuffel, and J.~A. Thompson, {\it {Calculation of the
  stopping-muon rate underground}},  {\em Phys. Rev.} {\bf D7} (1973) 2022.

\bibitem{Crouch:1987ICRC}
M.~{Crouch}, {\it {An Improved World Survey Expression for Cosmic Ray Vertical
  Intensity vs. Depth in Standard Rock}},  {\em International Cosmic Ray
  Conference} {\bf 6} (1987) 165.

\bibitem{Bohlen2014211}
T.~B{\"o}hlen, F.~Cerutti, M.~Chin, A.~Fass\`o, A.~Ferrari, P.~Ortega,
  A.~Mairani, P.~Sala, G.~Smirnov, and V.~Vlachoudis, {\it The {FLUKA} code:
  Developments and challenges for high energy and medical applications},  {\em
  Nuclear Data Sheets} {\bf 120} (2014) 211 -- 214.

\bibitem{Ferrari:2005zk}
A.~Ferrari, P.~R. Sala, A.~Fass\`o, and J.~Ranft, {\it {FLUKA: A multi-particle
  transport code}},  {\em CERN-2005-010, SLAC-R-773, INFN-TC-05-11} (2005).

\bibitem{Kuhlman_Malama_2013}
K.~L. Kuhlman and B.~Malama, {\it Brine flow in heated geologic salt.},  Tech.
  Rep. SAND2013-1944, Sandia National Laboratories, Mar, 2013.

\bibitem{Sorge:1989dy}
H.~Sorge, H.~Stoecker, and W.~Greiner, {\it {Poincare Invariant Hamiltonian
  Dynamics: Modeling Multi - Hadronic Interactions in a Phase Space Approach}},
   {\em Annals Phys.} {\bf 192} (1989) 266.

\bibitem{Roesler:2000he}
S.~Roesler, R.~Engel, and J.~Ranft, {\it {The Monte Carlo event generator
  DPMJET-III}},  \href{http://arxiv.org/abs/hep-ph/0012252}{{\tt
  hep-ph/0012252}}.

\bibitem{G4bugzilla}
Geant4, ``{Bugzilla/Geant4 - Problem 1680: photonuclear production cross
  section at giant dipole resonance wrong by factor 3}.''
\newblock {\url{http://bugzilla-geant4.kek.jp/show\_bug.cgi?id=1680}}.

\bibitem{Tilley:1993zz}
D.~R. Tilley, H.~R. Weller, and C.~M. Cheves, {\it {Energy levels of light
  nuclei A = 16-17}},  {\em Nucl. Phys.} {\bf A564} (1993) 1.

\bibitem{Browne20131849}
E.~Browne and J.~Tuli, {\it {Nuclear Data Sheets for A = 60}},  {\em Nuclear
  Data Sheets} {\bf 114} (2013) 1849.

\bibitem{SINGH200133}
B.~Singh, {\it {Nuclear Data Sheets for A = 130}},  {\em Nuclear Data Sheets}
  {\bf 93} (2001) 33.

\bibitem{Khazov2005497}
Y.~Khazov, A.~Rodionov, S.~Sakharov, and B.~Singh, {\it {Nuclear Data Sheets
  for A = 132}},  {\em Nuclear Data Sheets} {\bf 104} (2005) 497.

\bibitem{Sonzogni20041}
A.~Sonzogni, {\it {Nuclear Data Sheets for A = 134}},  {\em Nuclear Data
  Sheets} {\bf 103} (2004) 1.

\bibitem{Singh2008517}
B.~Singh, A.~A. Rodionov, and Y.~L. Khazov, {\it {Nuclear Data Sheets for A =
  135}},  {\em Nuclear Data Sheets} {\bf 109} (2008) 517.

\bibitem{mughabghab2006atlas}
S.~Mughabghab, {\em Atlas of Neutron Resonances: Resonance Parameters and
  Thermal Cross Sections. Z=1-100}.
\newblock Elsevier Science, 2006.

\bibitem{Prussin:1977zz}
S.~G. Prussin, R.~G. Lanier, G.~L. Struble, L.~G. Mann, and S.~M. Schoenung,
  {\it {Gamma rays from thermal neutron capture in Xe-136}},  {\em Phys. Rev.}
  {\bf C16} (1977) 1001.

\bibitem{Becvar1998434}
F.~Be\u{c}v\'{a}\u{r}, {\it Simulation of $\gamma$ cascades in complex nuclei
  with emphasis on assessment of uncertainties of cascade-related quantities},
  {\em Nucl. Instrum. Meth.} {\bf A417} (1998) 434.

\bibitem{Singh2007197}
B.~Singh, {\it {Nuclear Data Sheets for A = 64}},  {\em Nuclear Data Sheets}
  {\bf 108} (2007) 197. {Data extracted from the ENSDF database, version July
  7, 2013, \url{http://www.nndc.bnl.gov/}}.

\bibitem{Browne20101093}
E.~Browne and J.~K. Tuli, {\it {Nuclear Data Sheets for A = 66}},  {\em Nuclear
  Data Sheets} {\bf 111} (2010) 1093. {Data extracted from the ENSDF database,
  version July 7, 2013, \url{http://www.nndc.bnl.gov/}}.

\bibitem{tuliensdfformat}
J.~K. Tuli, {\it {Evaluated nuclear structure data file: A Manual for
  Preparation of Data Sets}},  Tech. Rep. BNL-NCS-51655-01/02 Rev., Brookhaven
  National Laboratory, 2001.

\bibitem{Tilley1998249}
D.~Tilley, C.~Cheves, J.~Kelley, S.~Raman, and H.~Weller, {\it {Energy levels
  of light nuclei, A = 20}},  {\em Nuclear Physics A} {\bf 636} (1998) 249.
  {Data extracted from the ENSDF database, version February 12, 2015,
  \url{http://www.nndc.bnl.gov/}}.

\bibitem{fluka134xe}
A.~Fass\`o, A.~Ferrari, P.~Sala, and G.~Tsiledakis, {\it {Implementation of
  Xenon Capture Gammas in FLUKA for TRD Background Calculations}},  in {\em
  Proceedings of the Sixth Meeting of the Task Force on Shielding Aspects of
  Accelerators, Targets and Irradiation Facilities}, pp.~397--406, NEA, Apr.,
  2002.

\bibitem{Wilks:1938dza}
S.~Wilks, {\it {The Large-Sample Distribution of the Likelihood Ratio for
  Testing Composite Hypotheses}},  {\em Annals Math.Statist.} {\bf 9} (1938)
  60.

\bibitem{Cowan:1998ji}
G.~Cowan, {\em {Statistical data analysis}}.
\newblock Oxford Science Publications. Clarendon, 1998.

\bibitem{Abe:2009aa}
{\bf KamLAND} Collaboration, S.~Abe et~al., {\it {Production of Radioactive
  Isotopes through Cosmic Muon Spallation in KamLAND}},  {\em Phys. Rev.} {\bf
  C81} (2010) 025807, [\href{http://arxiv.org/abs/0907.0066}{{\tt
  arXiv:0907.0066}}].

\bibitem{Reichhart:2013xkd}
L.~Reichhart et~al., {\it {Measurement and simulation of the muon-induced
  neutron yield in lead}},  {\em Astropart. Phys.} {\bf 47} (2013) 67,
  [\href{http://arxiv.org/abs/1302.4275}{{\tt arXiv:1302.4275}}].

\bibitem{Marino:2007ti}
M.~G. Marino, J.~A. Detwiler, R.~Henning, R.~A. Johnson, A.~G. Schubert, and
  J.~F. Wilkerson, {\it {Validation of spallation neutron production and
  propagation within Geant4}},  {\em Nucl. Instrum. Meth.} {\bf A582} (2007)
  611, [\href{http://arxiv.org/abs/0708.0848}{{\tt arXiv:0708.0848}}].

\bibitem{Bellini:2013pxa}
{\bf Borexino} Collaboration, G.~Bellini et~al., {\it {Cosmogenic Backgrounds
  in Borexino at 3800 m water-equivalent depth}},  {\em JCAP} {\bf 1308} (2013)
  049, [\href{http://arxiv.org/abs/1304.7381}{{\tt arXiv:1304.7381}}].

\bibitem{SOURCES}
W.~B. Wilson, {\it {SOURCES 4A : a code for calculating ($\alpha$,n)
  spontaneous fission, and delayed neutron sources and spectra}},  Tech. Rep.
  LA-13639-MS, Los Alamos National Laboratory, Los Alamos, NM, USA, 1999.

\bibitem{Firestone:391553}
R.~B. Firestone, C.~Baglin, and S.~Y.~F. Chu, {\em {Table of isotopes: 1998
  update with CD-ROM}}.
\newblock Wiley, New York, NY, 1998.

\end{thebibliography}\endgroup

\end{document}